\documentclass[aps,pre,superscriptaddress,twocolumn,10pt]{revtex4-2}

\usepackage[export]{adjustbox}
\usepackage{amsmath}
\usepackage{amssymb}
\usepackage{amsthm}
\usepackage{bbm}
\usepackage{color}
\usepackage{graphicx}[floatfix]
\usepackage{hyperref}
\usepackage[utf8]{inputenc}
\usepackage[T1]{fontenc}
\usepackage{orcidlink}
\usepackage{physics}
\usepackage{times}
\usepackage{ragged2e}
\usepackage[normalem]{ulem} 
\usepackage{soul,xcolor}
\usepackage{comment}

\hypersetup{
    colorlinks=true, linkcolor=blue, citecolor=blue,
    filecolor=blue, urlcolor=blue, breaklinks=true
}

\newcommand{\demat}{
  Departamento de Matem\'{a}tica e Estat\'{i}stica,
  Universidade Estadual de Ponta Grossa,
  84030-900 Ponta Grossa, Paraná, Brazil
}

\newcommand{\defis}{
  Departamento de Física,
  Universidade Estadual de Ponta Grossa,
  84030-900 Ponta Grossa, Paraná, Brazil
}

\newcommand{\usp}{
  Departamento de F\'{i}sica Matem\'{a}tica,
  Instituto de F\'{i}sica da Universidade de S\~{a}o Paulo,
  05508-090 S\~{a}o Paulo, Brazil
}

\newcommand{\qpqi}{
  QPQI Group,
  Universidade Estadual de Ponta Grossa,
  84030-900 Ponta Grossa, Paraná, Brazil
}

\newcommand{\orcidthiago}{\orcidlink{0009-0001-1654-0330}}
\newcommand{\orciddanilo}{\orcidlink{0000-0002-4177-1237}}
\newcommand{\orcidantonio}{\orcidlink{0000-0002-1521-9342}}
\newcommand{\orcidfabiano}{\orcidlink{0000-0001-5383-6168}}

\hypersetup{
    colorlinks=true,linkcolor=blue,citecolor=blue,
    filecolor=blue,urlcolor=blue,breaklinks=true
}

\RequirePackage{xcolor}

\newcommand{\doubt}[1]{{\color{magenta}{#1}}}

\begin{document}

\setstcolor{red}

\title{
Fractional-Time Jaynes-Cummings Model:  Unitary Description of its Quantum Dynamics, Inverse Problem and Photon Statistics
}

\author{Thiago T. Tsutsui\orcidthiago}
\email{takajitsutsui@gmail.com}
\affiliation{\qpqi}

\author{Danilo Cius\orciddanilo}
\email{danilocius@gmail.com}
\affiliation{\qpqi}
\affiliation{\usp}

\author{Antonio S. M. de Castro\orcidantonio}
\email{asmcastro@uepg.br}
\affiliation{\qpqi}
\affiliation{\defis}

\author{Fabiano M. Andrade\orcidfabiano}
\thanks{Corresponding author: \href{mailto:fmandrade@uepg.br}{fmandrade@uepg.br}}
\affiliation{\qpqi}
\affiliation{\demat}

\date{\today}

\begin{abstract}
We analyze the quantum dynamics of the fractional-time Jaynes-Cummings model using a recent unitary framework for the fractional-time Schr\"odinger equation. We examine how the fractional derivative order $\alpha$ influences non-classical features under different initial conditions. For an initial Fock state, fractional evolution introduces transient dynamics and heightened sensitivity to coupling strength. Through an inverse problem approach, we interpret these effects as arising from an effective time-dependent coupling with a strong initial pulse. For an initial coherent state, the fractional order tunes the system between dynamical regimes, with a transition at $\alpha = 0.50 $ where standard collapse-and-revival is replaced by stable, periodic evolution. This regime enhances non-classical field properties, including stronger sub-Poissonian statistics, periodic quadrature squeezing, and the formation of Schr\"odinger cat states.
\end{abstract}

\maketitle

\section{Introduction}
\label{sec:intro}

In the past few decades, there has been a growing theoretical interest in
extending the framework of standard quantum mechanics.
For example, fractional-order differential operators
\cite{podlubny1998fractional} have been used in quantum mechanics to
model nonlocal or anomalous effects, thanks to their success with classical
systems when integer-order methods do not work \cite{metzler2000random}.
Building up this idea, the broad applicability of the framework extends
to various fields of quantum mechanics, including optics
\cite{zhang2015,longhi2015fractional,zhang2016pt,huang2017beam,colas2020},
condensed matter \cite{wu2010,stickler2013,stephanovich2022,zhong2024},
and quantum information \cite{ zu2021,elallati2024,abdessamie2025,Gabrick2026}.
The concept of fractional quantum mechanics was independently introduced
by Laskin \cite{laskin:00a,laskin:00b,laskin2002} and West
\cite{west00}, who derived a fractional-space Schrödinger equation
(FSSE) through Feynman's path integral formulation \cite{feynman1965},
where the integration is performed not only over Brownian trajectories
but over paths exhibiting Lévy flights.
In these works, only the spatial derivative is extended to a fractional
order, meaning that the order of differentiation is not necessarily an
integer but a real number between 0 and 1.
The time derivative, however, remains of the usual first order.
An interesting application of FSSE appears, for example, in quantum
transport in networks \cite{riascos2015}, where a long-range quantum
walk is introduced via FSSE to study the efficiency of quantum
transport.
Physical implementations of the FSSE are discussed in
Refs. \cite{stickler2013,longhi2015fractional,zhang2017}.

Inspired by Laskin’s earlier work, Naber \cite{naber:04} proposed the
fractional-time Schr\"odinger equation (FTSE) by transforming the
fractional Fokker-Planck equation into a Schr\"{o}dinger-like form
through a Wick rotation of time,  $t \to -it/\hbar$.
In his formulation, the ordinary time derivative is replaced by the
Caputo fractional derivative, which causes the imaginary unit to appear
raised to the same fractional order as the time derivative.
Then, the FTSE is written as
\begin{equation}
  \label{FTSE}
  i^{\alpha}{\,}^{\text{C}}_0\mathcal{D}_t^{\alpha} \ket{\Psi_{\alpha}(t)}
  = \hat{H}_\alpha \ket{\Psi_{\alpha}(t)},
\end{equation}
where $\hat{H}_\alpha$ is the fractional Hamiltonian operator, while
${}^{~\text{C}}_{~0}\mathcal{D}_t^{\alpha}$ represents the Caputo
derivative operator, which is given by
\begin{equation}
  \label{FD}
  {}^{~\text{C}}_{~0}\mathcal{D}_t^{\alpha}(\cdot)
  = \frac{1}{\Gamma(1-\alpha)~}
  \int_{0}^{t}d\tau(t-\tau)^{-\alpha}\frac{d }{d\tau}(\cdot),
\end{equation}
when assuming $\alpha \in (0,1]$, and $\Gamma(\cdot)$ denotes the
well-known Gamma function.
It should be noted that the standard Schrödinger equation is recovered
in the limit $\alpha \to 1$.
\emph{Here, we consider all variables to be dimensionless}.
Solutions to FTSE have been investigated in various contexts, including
the fractional dynamics of free particles \cite{naber:04}, systems under
the influence of  $\delta$-function potentials \cite{lenzi2013time},
time-dependent quantum potentials \cite{gabrick2023}, and fractional
dynamics of the comb model \cite{iomin2009fractional,iomin2024}.
While the FSSE describes Markovian processes, the FTSE accounts for
non-Markovian evolution with memory effects and dissipation
\cite{naber:04,Zu2025}. Such memory-dependent quantum dynamics typically
arises from system-environment interactions, leading to non-Markovian
and non-unitary behavior \cite{breuer2007,tarasov2010,tarasov2012}.
Furthermore, generalizations of the fractional Schr\"odinger equation in
space and time have been reported in \cite{wang2007,dong2008}, and their
application to a three-level system demonstrates anomalous relaxation
and wave packet spreading \cite{lenzi2023}.

Fractional quantum mechanics introduces intriguing modifications to
standard quantum theory.
While the fractional-space derivative preserves the principles of
quantum mechanics, the fractional-time derivative violates the
fundamental laws of quantum mechanics \cite{laskin2017time}.
For instance, the Caputo derivative leads to non-unitary dynamics,
resulting in non-conservation of probability due to an effective
non-Hermitian Hamiltonian
\cite{ertik2010,laskin2017time,iomin2019fractional}.
This also breaks Stone’s Theorem \cite{stone1932}, which establishes a
link between strongly continuous one-parameter unitary groups and
self-adjoint operators in a Hilbert space.
Despite these difficulties, progress has been made toward restoring
unitary dynamics in the fractional-time framework.
Recently, a consistent unitary formulation of the FTSE was proposed in
\cite{cius22frac}, which preserves probability while retaining the
essential features of fractional dynamics.
This development paves the way for applying fractional-time quantum
mechanics to well-established models in quantum optics and quantum
information, where unitarity is crucial for preserving physical
consistency.

The Jaynes-Cummings (JC) model is a foundational model in cavity quantum
electrodynamics (QED) \cite{jaynes1963}, which describes the interaction
between a two-level atom and a single quantized field mode, and it is a
standard system for studying characteristic quantum effects, such as
Rabi oscillations (RO), collapse and revival phenomena, and the
generation of entanglement and nonclassical light states.
Applying the unitary framework for FTSE to the JC model, therefore,
provides a method to investigate how fractional-time derivatives modify
these fundamental quantum optical processes.
Furthermore, even though the JC model has already been extensively
studied, recent work continues to reveal new aspects of its dynamics by
incorporating different physical features
\cite{UHDRE2022,VidiellaBarranco2025,Arrazola2025,Zhou2025a,Tsutsui2026}.

The unitary description of the fractional-time JC model (FTJC) was
recently introduced in \cite{Cius2024}, where the analysis focuses on
the population inversion dynamics and atom-field entanglement for an
atom prepared in the excited state and a cavity in a coherent state.
Our work extends and complements this investigation by exploring
different physical regimes.
We provide a detailed analysis of the atomic dynamics for the initial
Fock states and a comprehensive study of the photon statistics for
initial coherent states, aspects not previously considered in this
fractional-time context.
For initial Fock states, we observe transient effects in the first two
oscillations of the atomic population and entanglement, as well as a
heightened dependence on the coupling strength.
Using an inverse problem approach (IPA), we show that time-dependent
couplings with a strong pulse at the outset can replicate
fractional-time effects.

This paper is organized as follows. In Sec. \ref{sec:jc_model}, we
briefly introduce the JC model. Building on this, we establish the
fractional-time description of the model and its unitary counterpart.
Our main results are presented in the following sections.
In Sec. \ref{sec:fock_frac_jc_model}, we analyze the dynamics of the
system for an initial Fock state, focusing on the atomic inversion and
atom-field entanglement to identify how the fractional order $\alpha$
modifies the RO.
In Sec. \ref{sec:photon}, we investigate the photon statistics for an
initial coherent state, examining the effects on sub-Poissonian
behavior, squeezing, and the generation of Schrödinger cat states.
We show that for $\alpha = 0.50$, the system exhibits periodic squeezing
and cat-state formation, the latter being a feature of the two-photon JC
model \cite{Puri1988}.
Finally, Sec. \ref{sec:Conc} summarizes our conclusions and discusses
the broader implications of our findings in light-matter interactions.

\section{Unitary Framework for the fractional-time Jaynes-Cummings Model}
\label{sec:jc_model}

Recently, an extension of the JC model to the FTSE context was proposed, employing non-Hermitian quantum formalism to map a non-unitary time
evolution to a unitary one \cite{Cius2024}.
In that instance, the dynamics were investigated with the cavity mode
initially in a coherent state, focusing on population inversion and
atom-field entanglement as quantified by the von Neumann entropy.
In the standard JC model ($\alpha=1.00$), this condition leads to
collapses and revivals of the RO, a signature of the quantum nature of
the light field.
However, the introduction of fractional-time evolution altered the
behavior of the atomic probabilities.
In particular, when $\alpha=0.50$, the characteristic collapses and
revivals were replaced by persistent, periodic oscillations, resembling
the dynamics of a driven system.
Additionally, modifications in the evolution of the atom-field entanglement
were observed.

Below, we summarize the unitary framework established in
Ref. \cite{Cius2024}, which provides the basis for a consistent physical
analysis.
We begin by reviewing the JC Hamiltonian, then detailing the non-unitary
fractional evolution and the Dyson map formalism applied to restore a
probabilistic interpretation.

\subsection{Jaynes-Cummings model}

The JC model \cite{jaynes1963} stands as a fundamental framework for
describing light-matter interaction, depicting the dynamics of a
two-level atom coupled to a quantized cavity mode.
Derived from the Rabi model \cite{Rabi1936,Rabi1937} under the
rotating-wave approximation,
the JC Hamiltonian is given by
\begin{equation}
  \label{eq:JC_Hamilt_full}
    \hat{H}_\alpha =
    \frac{1}{2} \omega_\alpha \hat{\sigma}^z
    + \nu_\alpha \hat{a}^\dagger \hat{a}
    + \mu_\alpha ( \hat{\sigma}^{+}\hat{a} +
    \hat{\sigma}^{-}\hat{a}^{\dagger}),
\end{equation}
where $\omega_\alpha$ is the transition frequency of the atom, $\nu_\alpha$
is the cavity mode frequency, and $\mu_\alpha$ is the atom-field
coupling strength.
The atom is described by the Pauli operators  $\hat{\sigma}^z$ and
$\hat{\sigma}^{\pm}$, which form an $\mathfrak{su}(2)$ algebra
characterized  by the commutation relation
$[\hat{\sigma}^{+},\hat{\sigma}^{-}] =  2\hat{\sigma}^{z}$ \cite{Klimov2009}.
On the other hand, the degree of freedom of the cavity mode is represented  by
the creation and annihilation operators, $\hat{a}^\dagger$ and
$\hat{a}$, respectively, elements of the Weyl-Heisenberg algebra
$[\hat{a},\hat{a}^{\dagger}] =  \hat{1}$ \cite{Cantuba2024}.
Assuming resonance ($\omega_\alpha = \nu_\alpha$) and employing the
interaction picture, the Hamiltonian, Eq. \eqref{eq:JC_Hamilt_full},
simplifies to
\begin{equation}
\label{eq:JC_Hamilt_int_pic}
    \hat{V}_\alpha =  \mu_\alpha (\hat{\sigma}^{+}\hat{a} +
    \hat{\sigma}^{-}\hat{a}^{\dagger}).
\end{equation}
Physically, the term $\hat{\sigma}^{+}\hat{a}$ represents the absorption
of a cavity photon, causing the atom to transition from the ground to
the excited state, while the term $\hat{\sigma}^{-}\hat{a}^\dagger$
represents the emission of a photon into the cavity, as the atom decays
from the excited state to the ground state.

Additionally, the JC Hamiltonian commutes with the excitation number
operator $\hat{N}_E=\hat{a}^\dagger \hat{a} + \hat{\sigma}^z/2$, which
allows the Hilbert space to be decomposed into a direct sum of
orthogonal subspaces,
$\mathcal{H} = \mathcal{H}_{\text{ground}} \oplus \left(
  \bigoplus_{n=0}^\infty \mathcal{H}_n \right)$.
Here, $\mathcal{H}_{\text{ground}}$ is the one-dimensional subspace
spanned by the vacuum state $|g,0\rangle$, and each $\mathcal{H}_n$ is
spanned by the basis $\{|e,n\rangle, |g,n+1\rangle \}$
\cite{larson2021}.
Consequently, the Hamiltonian itself can be expressed as a direct sum of
operators $\hat{V}^{(n)}$ that act only within these subspaces,
$\hat{V}_\alpha = \bigoplus_{n=0}^\infty \hat{V}_{\alpha}^{(n)}$.
This structure, when written in matrix form, is block-diagonal, where
each block is a $2 \times 2$ matrix representing the Hamiltonian in  the
subspace $\mathcal{H}_n$:
\begin{equation}
  \label{eq:H0_block_matrix}
  \hat{V}_{\alpha}^{(n)} = \mu^{(n)}_{\alpha}
  \begin{pmatrix}
    0 & 1 \\
    1 & 0
  \end{pmatrix},
\end{equation}
where $\mu^{(n)}_{\alpha} = \mu_\alpha\sqrt{n+1}$ is the effective
photon-number-dependent coupling strength.

\subsection{Non-Unitary Fractional-Time Evolution}
The fractional-time framework involves replacing the standard Schrödinger equation with the FTSE. The formal solution of Eq. \eqref{FTSE} is given by $|\Psi_{\alpha}(t)\rangle = \hat{U}_{\alpha}(t) |\Psi_{\alpha}(0)\rangle$, where the time-evolution operator $\hat{U}_{\alpha}(t)$ assumes the following form
\begin{eqnarray}
   \hat{U}_{\alpha}(t) = E_{\alpha}\left(i^{-\alpha}\hat{V}_{\alpha}\,t^{\alpha}\right),
\end{eqnarray}
satisfying the initial condition $\hat{U}_{\alpha}(0)=\hat{1}$.
In the above equation, the function  $E_{\alpha}(x)= \sum_{k=0}^{\infty} x^{k}/\Gamma(\alpha k +1)$  is identified to be the well-known one-parameter Mittag-Leffler function \cite{podlubny1998fractional}.
This formulation leads to a non-unitary evolution, which can be understood by mapping the fractional equation to a standard Schrödinger equation governed by an effective time-dependent non-Hermitian Hamiltonian \cite{laskin2017time,iomin2019app, cius22frac}.
As a result, the norm of the state vector is not conserved, which precludes a consistent probabilistic interpretation within a standard Hilbert space.

The time-evolution operator, being a function of the Hamiltonian, inherits its direct sum structure: $\hat{U}_{\alpha}(t) = \bigoplus_{n=0}^\infty \hat{U}_{\alpha}^{(n)}(t)$.
Each $\hat{U}_{\alpha}^{(n)}(t)$ acts independently within the subspace $\mathcal{H}_n$.
The matrix representation for each block is given by
\begin{equation}
\label{eq:U_matrix_frac}
\hat{U}_{\alpha}^{(n)}(t) =
\begin{pmatrix}
\mathcal{C}^{(n)}_{\alpha}(t) & i^{-\alpha}\mathcal{S}^{(n)}_{\alpha}(t) \\
i^{-\alpha}\mathcal{S}^{(n)}_{\alpha}(t) & \mathcal{C}^{(n)}_{\alpha}(t)
\end{pmatrix}.
\end{equation}
The complex functions $\mathcal{C}^{(n)}_{\alpha}(t)$ and $\mathcal{S}^{(n)}_{\alpha}(t)$ are defined as:
\begin{subequations}
\begin{align}
    \mathcal{C}^{(n)}_{\alpha}(t) &= \frac{E_{\alpha}(i^{-\alpha}\mu^{(n)}_{\alpha}t^{\alpha}) + E_{\alpha}(-i^{-\alpha}\mu^{(n)}_{\alpha}t^{\alpha})}{2}, \\
    \mathcal{S}^{(n)}_{\alpha}(t) &= \frac{E_{\alpha}(i^{-\alpha}\mu^{(n)}_{\alpha}t^{\alpha}) - E_{\alpha}(-i^{-\alpha}\mu^{(n)}_{\alpha}t^{\alpha})}{2i^{-\alpha}}.
\end{align}
\end{subequations}
The operator $\hat{U}_{\alpha}^{(n)}(t)$ is non-unitary. Consequently, it requires a formalism that preserves probability conservation and ensures a consistent physical interpretation.

\subsection{Restoring Unitarity via a Dyson Map}
\begin{figure}[t!]
  \centering
  \includegraphics[width=1\linewidth]{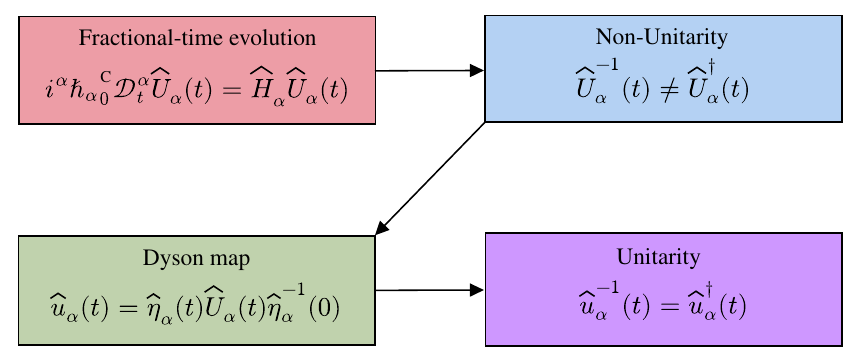}
  \caption{
   A diagram for the technique employed in this work.
   The fractional-time evolution leads to a non-unitary evolution.
   To address this, we employ a time-dependent Dyson map to obtain the
   unitary operator governing the evolution of the system.
  }
  \label{fig:scheme}
\end{figure}

To restore a consistent probabilistic framework, we employ a time-dependent Dyson map $\hat{\eta}_{\alpha}(t)$, which is an invertible operator that relates the non-unitary and unitary pictures \cite{fring:16a,fring:17,cius22frac}.
The physical state $|\psi_{\alpha}(t)\rangle = \hat{\eta}_{\alpha}(t)|\Psi_{\alpha}(t)\rangle$ evolves unitarily according to the operator
\begin{equation}
\label{eq:unitaryU}
    \hat{u}_{\alpha}(t) = \hat{\eta}_{\alpha}(t)\,\hat{U}_{\alpha}(t)\,\hat{\eta}_{\alpha}^{-1}(0),
\end{equation}
where $\hat{u}^{-1}_{\alpha}(t)=\hat{u}^{\dagger}_{\alpha}(t)$.
This ensures probability conservation through a modified, time-dependent inner product, defined by the metric operator $\hat{\Theta}_{\alpha}(t) = \hat{\eta}_{\alpha}^{\dagger}(t) \hat{\eta}_{\alpha}(t)$,
such that $\langle \Psi_{\alpha}(t) | \hat{\Theta}_{\alpha}(t) | \Psi_{\alpha}(t)\rangle = \langle \psi_{\alpha}(t) | \psi_{\alpha}(t)\rangle = 1$.
Our approach is schematically synthesized in Fig. \ref{fig:scheme}.

The Dyson map is constructed in a block-diagonal form,
$\hat{\eta}_{\alpha}(t) = \bigoplus_{n=0}^\infty
\hat{\eta}_{\alpha}^{(n)}(t)$.
For each subspace $\mathcal{H}_n$, we choose a Hermitian map represented
by the matrix:
\begin{equation}
\label{eq:etamat}
\hat{\eta}_{\alpha}^{(n)}(t) = \frac{e^{\kappa_{\alpha}^{(n)}(t)}}{\sqrt{\Lambda_{\alpha}^{(n)}(t)}}
\begin{pmatrix}
\chi_{\alpha}^{(n)}(t) & \lambda_{\alpha}^{(n)}(t) \\
[\lambda_{\alpha}^{(n)}(t)]^{\ast} & 1
\end{pmatrix},
\end{equation}
parametrized by the time-dependent real functions
$\kappa_{\alpha}^{(n)}(t)$ and $\Lambda_{\alpha}^{(n)}(t) > 0$, the
complex function $\lambda_{\alpha}^{(n)}(t)$, and
$\chi_{\alpha}^{(n)}(t)=\Lambda_{\alpha}^{(n)}(t) +
|\lambda_{\alpha}^{(n)}(t)|^{2}$. 

For $\hat{u}_{\alpha}^{(n)}(t)$ to be unitary, the parameters of the Dyson map must satisfy a set of specific conditions derived from the relation in Eq. \eqref{eq:unitaryU}.
These conditions, detailed in Refs. \cite{cius22frac,Cius2024}, ensure
that the resulting operator for each subspace,
$\hat{u}_{\alpha}^{(n)}(t)$, belongs to the $U(2)$ group and takes the
general form:
\begin{equation}
\label{eq:u_matrix_final}
\hat{u}^{(n)}_{\alpha}(t) = e^{i\delta_{\alpha}^{(n)}(t)}
\begin{pmatrix}
\varpi_{\alpha,+}^{(n)}(t) & \varpi_{\alpha,-}^{(n)}(t) \\
-[\varpi_{\alpha,-}^{(n)}(t)]^{\ast} & [\varpi_{\alpha,+}^{(n)}(t)]^{\ast}
\end{pmatrix},
\end{equation}
where $\delta_{\alpha}^{(n)}(t)$ is a time-dependent overall phase, the complex functions $\varpi_{\alpha,\pm}^{(n)}(t)$ determine the evolution amplitudes within the subspace $\mathcal{H}_n$, and the condition $|\varpi_{\alpha,+}^{(n)}|^{2} + |\varpi_{\alpha,-}^{(n)}|^{2} = 1$ is satisfied at all times.
To ensure this unitary structure, the parameters of the Dyson map must take a specific form determined by both the non-unitary evolution functions $\mathcal{C}_\alpha^{(n)}$ and $\mathcal{S}_\alpha^{(n)}$ as well as by the initial values of the map at $t=0$.
We set $\hat{\eta}_{\alpha}(0)=\hat{1}$, ensuring that the state vectors in the unitary and non-unitary pictures coincide at $t=0$: $\vert\psi_\alpha(0)\rangle = \vert\Psi_\alpha(0)\rangle$.
The explicit analytical solutions for these parameters are provided in Appendix \ref{app:solutions}, and the complete derivation is available in Refs. \cite{cius22frac,Cius2024}.

Therefore, the unitary description of the FTJC model is mathematically consistent and physically interpretable.
The essential features of fractional-time dynamics are preserved, such as anomalous modifications of RO, while ensuring probability conservation. This framework provides the foundation for analyzing population inversion, entanglement, and photon statistics in the following sections.

\section{Dynamics with an Initial Fock State}
\label{sec:fock_frac_jc_model}
The evolution of the JC system from an initial Fock state is a foundational scenario for studying the quantum nature of the light-matter interaction \cite{GERRY2005}.
The significance of this case extends to broader contexts,
as the single-photon case is an appropriate framework for the analysis of quantum information in cavity QED \cite{Nielsen2010}, making the control of Fock states a key practical objective \cite{Zhang2024}.
Therefore, in this section, we analyze the FTJC model under these conditions, focusing on three aspects: atomic population inversion, entanglement dynamics, and a method for simulating fractional evolution within a standard JC framework.

The evolution of the physical state is governed by the unitary operator $\hat{u}_{\alpha}(t)$ given by Eq. \eqref{eq:u_matrix_final}.
When the system is initialized with the atom in its excited state, the
state of the system is $\vert\psi_\alpha(t)\rangle$, for any time,
\begin{equation} \label{eq:state_JC}
  \vert\psi_\alpha(t)\rangle=
  \sum_{n=0}^\infty \left[ A_{e,n}^\alpha (t) |e,n\rangle +
    A_{g,n}^\alpha (t) |g,n+1\rangle \right],
\end{equation}
where $A_{e,n}^\alpha (t)$ and $A_{g,n+1}^\alpha (t)$ are probability amplitudes.
The probabilities associated with the excited and ground states are given, respectively, by the sums over the field states:
\begin{subequations}
\begin{align}
    P_e^\alpha (t) &=\sum_{n=0}^\infty|A_{e,n}^\alpha (t)|^2, \\
     P_g^\alpha (t)&=\sum_{n=0}^\infty|A_{g,n}^\alpha (t)|^2.
\end{align}
\end{subequations}

The study of the JC model is usually centered on population inversion $W_\alpha (t) = \langle \hat{\sigma}^z (t)\rangle$,
which is expressed in terms of atomic probabilities as
\begin{equation} \label{eq:pop_inversion}
    W_\alpha (t) = P_e^\alpha(t)- P_g^\alpha(t).
\end{equation}
This quantity is experimentally accessible \cite{REMPE1987,Brune1996} and provides a sensitive probe to analyze extensions of the JC model \cite{Arroyo-Correa1990}.
We begin our investigation with a detailed study of this observable.

\subsection{Population Inversion}

\begin{figure}[t]
  \centering
  \includegraphics[width=1\linewidth]{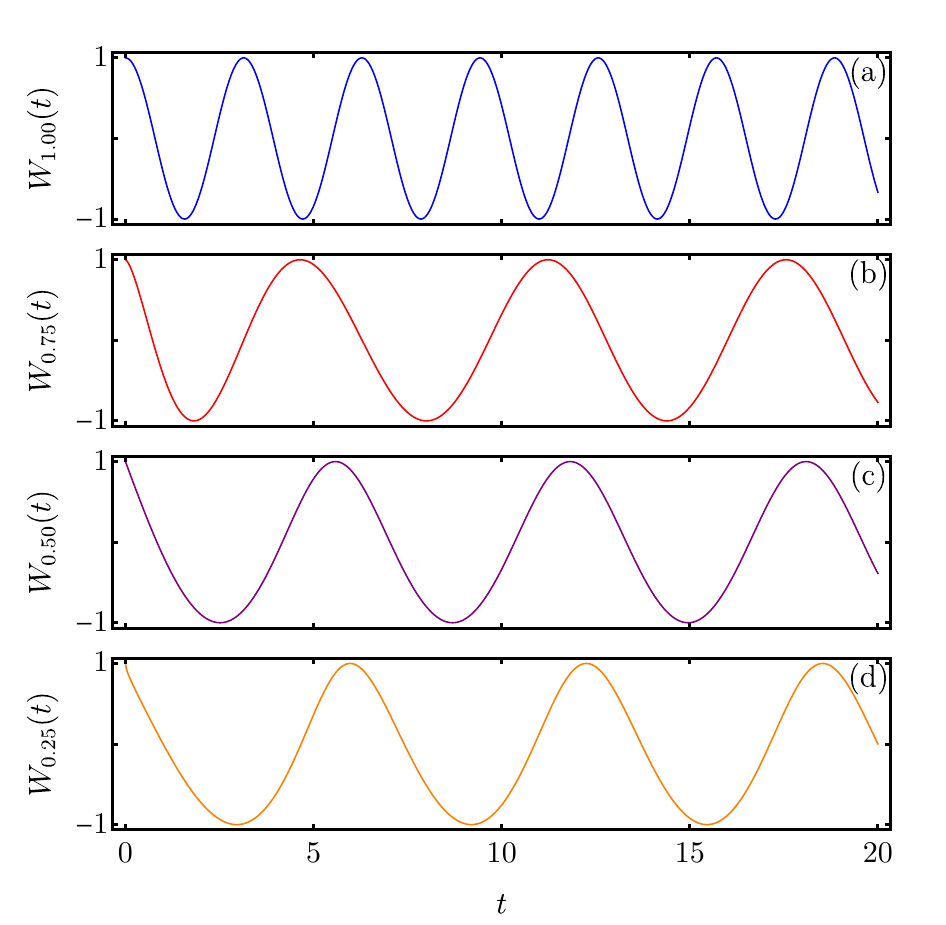}
  \caption{
    The population inversion $W_\alpha(t)$, Eq. \eqref{eq:pop_inversion},
    for the initial state $\ket{\Psi_\alpha(t)}=\ket{e,0}$ and
    $\mu_\alpha=1$, for different values of $\alpha$:
    (a) $\alpha=1.00$, (b) $\alpha=0.75$, (c) $\alpha=0.50$, and
    (d) $\alpha=0.25$.
    The initial values of the Dyson map used are
    $\kappa_\alpha(0)=\lambda_\alpha(0)=0$ and $\Lambda_\alpha(0)=1$.
    When $\alpha < 1.00$, we observe RO with longer periods and, in the
    first moments, effectively subtle aperiodicity.
  }
  \label{fig:pop_inv}
\end{figure}

\begin{figure}[b]
  \centering
  \includegraphics[width=1\linewidth]{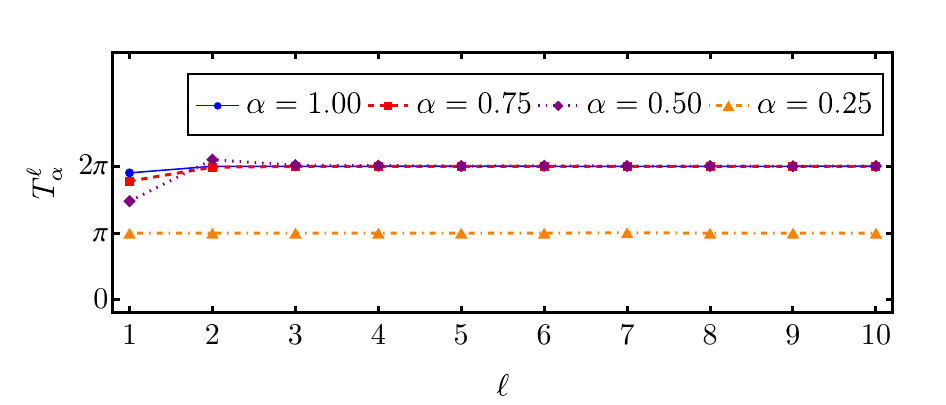}
  \caption{
    The periods $T_\alpha^\ell$ of the first ten oscillations in the
    population inversion $W_\alpha(t)$, considering the initial state
    $\ket{\psi_\alpha(0)}=\ket{e,0}$ and $\mu_\alpha=1$, for different
    values of $\alpha$:
    $\alpha=1.00$ (blue solid line with circles),
    $\alpha=0.75$ (red dashed line with squares),
    $\alpha=0.50$ (purple dotted line with diamonds), and
    $\alpha=0.25$ (orange dot-dashed line with triangles).
    The initial values of the Dyson map used are
    $\kappa_\alpha(0)=\lambda_\alpha(0)=0$ and $\Lambda_\alpha(0)=1$.
    After the first two oscillations, a constant period is observed for
    $\alpha < 1.00$.
  }
\label{fig:periods}
\end{figure}

When we consider the initial state $|\psi_\alpha(0)\rangle=|e,0\rangle$ in the FTJC,
the resulting population  inversion, $W_{\alpha}(t)$, is shown in Fig. \ref{fig:pop_inv}.
For the standard model ($\alpha=1.00$), the dynamics corresponds to the well-known vacuum RO, i.e., $W_{1.00}(t)=\cos(2 \mu_{1.00} t)$.
In contrast, for fractional cases with $\alpha=0.75$, $\alpha=0.50$, and $\alpha=0.25$, the oscillation periods are longer, and the dynamics display a transient aperiodic behavior for the first few oscillations.
To approach this matter, in Fig. \ref{fig:periods}, we present the period $T_\alpha^{\ell}$ of the $\ell$-th oscillation, for the different $\alpha$ considered.

We observe that the first two oscillations, for $\alpha \neq 1.00$,
exhibit transient effects, before the period stabilizes at $2 \pi$.
This suggests that the effects of fractional-time derivatives on the
dynamics can be interpreted as those of an external field, whose
influence is more pronounced at early times.
An additional noteworthy aspect of our model is that the probabilistic
behavior is distinct from that observed in other fractional JC models,
such as those based on different fractional derivatives \cite{Wei2024}.

\begin{figure}
  \includegraphics[width=1\linewidth]{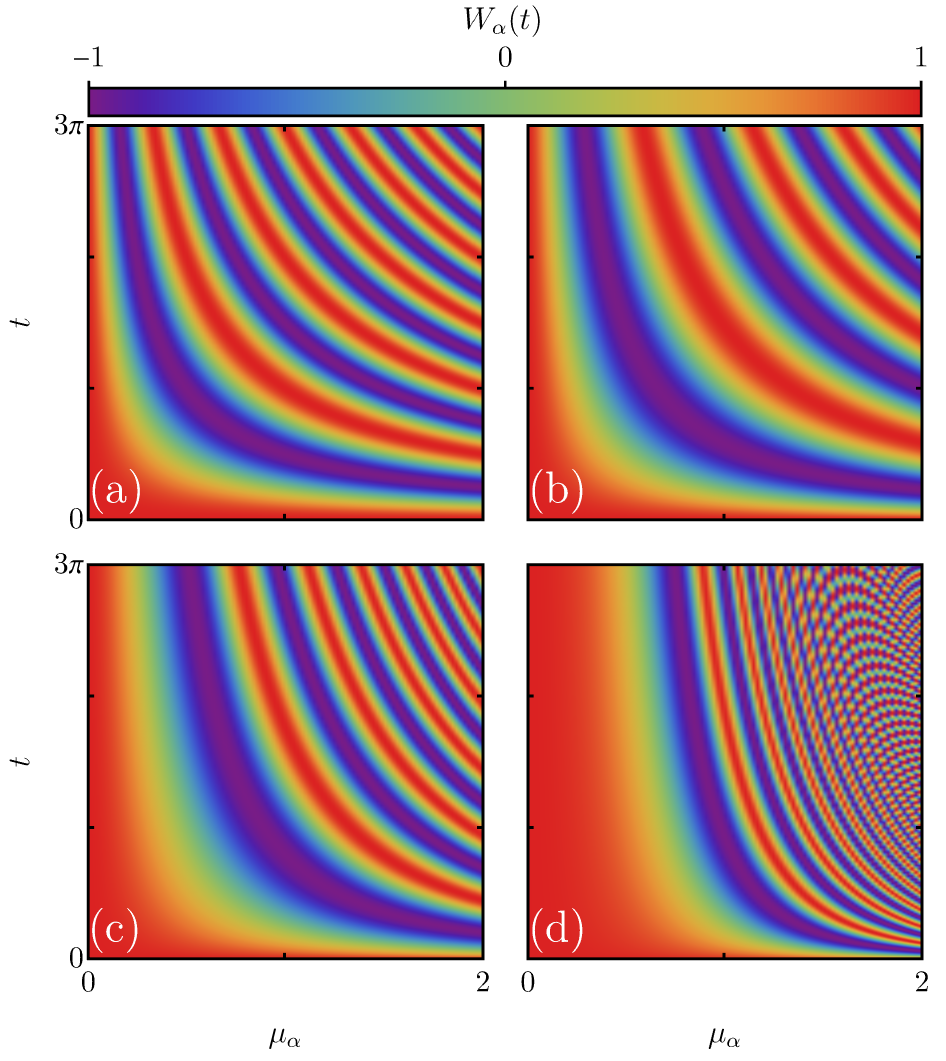}
  \caption{
    The effects of different coupling parameters on the population
    inversion (color bar), considering the initial state
    $\ket{\psi_\alpha(0)}=\ket{e,0}$, for different values of $\alpha$:
    (a) $\alpha=1.00$, (b) $\alpha=0.75$, (c) $\alpha=0.50$, and
    (d) $\alpha=0.25$.
    The initial values of the Dyson map used are
    $\kappa_\alpha(0)=\lambda_\alpha(0)=0$ and $\Lambda_\alpha(0)=1$.
    For $\alpha \leq 0.5$, the population inversion exhibits a greater
    sensitivity to the coupling parameter.
  }
  \label{fig:griddensity}
\end{figure}

In Fig. \ref{fig:griddensity}, we investigate the influence of the
coupling strength $\mu_\alpha$.
In the standard scenario, the number of oscillations within a fixed time
window is directly proportional to the coupling parameter.
For example, at the instants $t = 3 \pi$ and $\mu_\alpha = 1$, three
oscillations have already occurred, as shown in
Fig. \ref{fig:griddensity}(a).
Under the same conditions, assuming $\mu_\alpha = 2$, the number of
oscillations doubles.
However, when $\alpha \leq 0.50$, this proportionality becomes more
sensitive to the coupling strength.
This increased sensitivity is particularly remarkable in
Fig. \ref{fig:griddensity}(d), which shows that for $\mu_{\alpha} = 2$,
the system undergoes a significantly greater number of oscillations than
in the standard case.
This suggests that the fractional-time evolution introduces an effective
non-linearity into the system response to the interaction strength.

\subsection{Concurrence}

\begin{figure}
  \centering
  \includegraphics[width=1\linewidth]{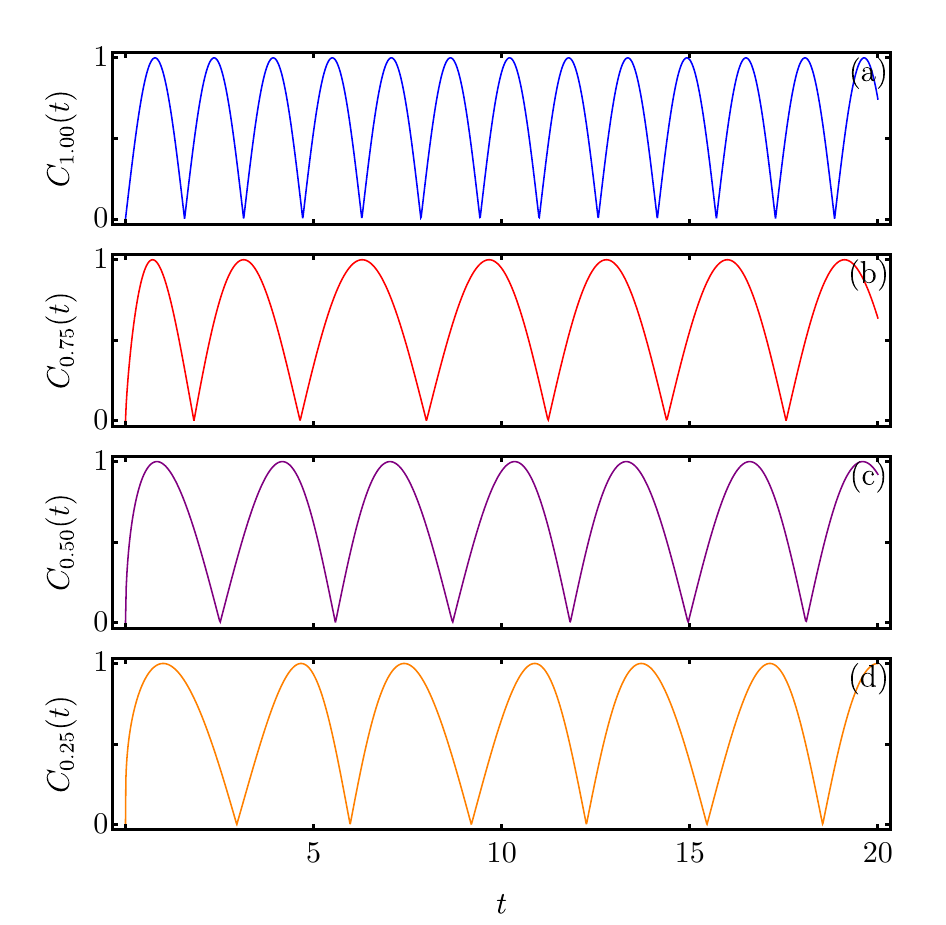}
  \caption{
  The concurrence $C_\alpha(t)$, considering the initial state $|\psi_\alpha(0)\rangle=|e,0\rangle$ and $\mu_\alpha=1$, for different values of $\alpha$: (a) $\alpha=1.00$, (b) $\alpha=0.75$, (c) $\alpha=0.50$,  and (d) $\alpha=0.25$ .
  The initial values of the Dyson map used are $\kappa_\alpha(0)=\lambda_\alpha(0)=0$ and $\Lambda_\alpha(0)=1$.
  Variations in the period and transient effects are noticeable for $\alpha < 1.00$.
 }
  \label{fig:conc}
\end{figure}

To quantify the atom-field entanglement in the one-photon scenario, we employ the concurrence \cite{Wootters1998}, defined as
\begin{equation}
    C_\alpha (t) = \max[0, \iota^1_\alpha(t) - \iota^2_\alpha(t) - \iota^3_\alpha(t) - \iota^4_\alpha(t)],
\end{equation}
where $\iota^k_\alpha(t)$ are the eigenvalues, in decreasing order, of the matrix
$
    R_\alpha (t) = \sqrt{\sqrt{\hat{\rho}_\alpha (t)} \Tilde{\hat{\rho}}_\alpha (t)\sqrt{\hat{\rho}_\alpha (t)}},
$
with $\hat{\rho}_\alpha (t) =|\psi_\alpha (t)\rangle\langle\psi_\alpha (t)|$,
and $\Tilde{\hat{\rho}}_\alpha (t)=(\hat{\sigma}_y\otimes\hat{\sigma}_y)
\hat{\rho}_\alpha(t)(\hat{\sigma}_y\otimes\hat{\sigma}_y)$.
The evolution of the concurrence in the FTJC is shown in Fig. \ref{fig:conc}.
For the initial state considered, the behavior of the concurrence can be understood from the population inversion: when $|W_\alpha(t)|=1$, the entanglement is zero because the system surely is in $|e,n\rangle$ or in $|g,n+1\rangle$.
However, if $W_\alpha(t)=0$, the probabilities associated with each of the bare states are equal, and the entanglement reaches the maximum value.
Thus, for $\alpha=1.00$, the concurrence exhibits similar periodicity to
the standard RO [Fig. \ref{fig:conc}(a)]. 
Similarly, the entanglement dynamics for $\alpha<1.00$ exhibit the same transient behavior and prolonged periods found in the population inversion [see Figs. \ref{fig:conc}(b), (c), and (d)].

\subsection{Inverse Problem Approach}
To provide a more intuitive physical interpretation and a potential route for experimental simulation, we employ the IPA.
Within a standard quantum-mechanical framework, the objective is to determine the time-dependent coupling, denoted $\gamma_{\alpha}(t)$, that reproduces the population dynamics $W_{\alpha}(t)$ generated by our fractional model.
This approach is particularly relevant, as \emph{in situ} access to
coupling strengths is experimentally feasible in circuit QED systems
\cite{Gambetta2011,Srinivasan2011,Yin2013,Srinivasan2014,Zeytinoglu2015},
which provide a platform for the JC model.

For a standard JC model initialized in the state $\vert e, 0\rangle $, the time-dependent coupling required to produce a target population inversion, $W_{\alpha}(t)$, can be derived using the IPA formalism \cite{Yang2006}.
The general solution for the coupling, $\gamma_{\alpha}(t)$, is a complex function whose magnitude depends on $W_\alpha(t)$. However, as we demonstrate in Appendix \ref{app:ipa}, the population inversion dynamics are determined solely by the magnitude of this coupling $|\gamma_{\alpha}(t)|$.
The phase of the coupling affects only the relative phase of the quantum state amplitudes, which is not directly observable in the population.

Therefore, for the purpose of reproducing a target $W_{\alpha}(t)$, we are free to select a real, strictly positive coupling strength defined by the following expression
\begin{equation} \label{eq:ipa_formula}
    \gamma_\alpha (t) = \left\vert \frac{\dot{W}_\alpha(t)}{2\sqrt{1-W_\alpha^2(t)}}\right\vert .
\end{equation}
We employ another symbol for the coupling in this context  to symbolize that this coupling acts on the non-fractional JC model,
where we investigate which time-dependent coupling can reproduce the effects of fractional-time on the population inversion.
Here, the population inversion $W_\alpha (t)$ obtained from our fractional model serves as input to determine the control function $\gamma_\alpha(t)$ that would be needed to mimic the dynamics in a conventional experiment.

\begin{figure}
  \centering
  \includegraphics[width=1\linewidth]{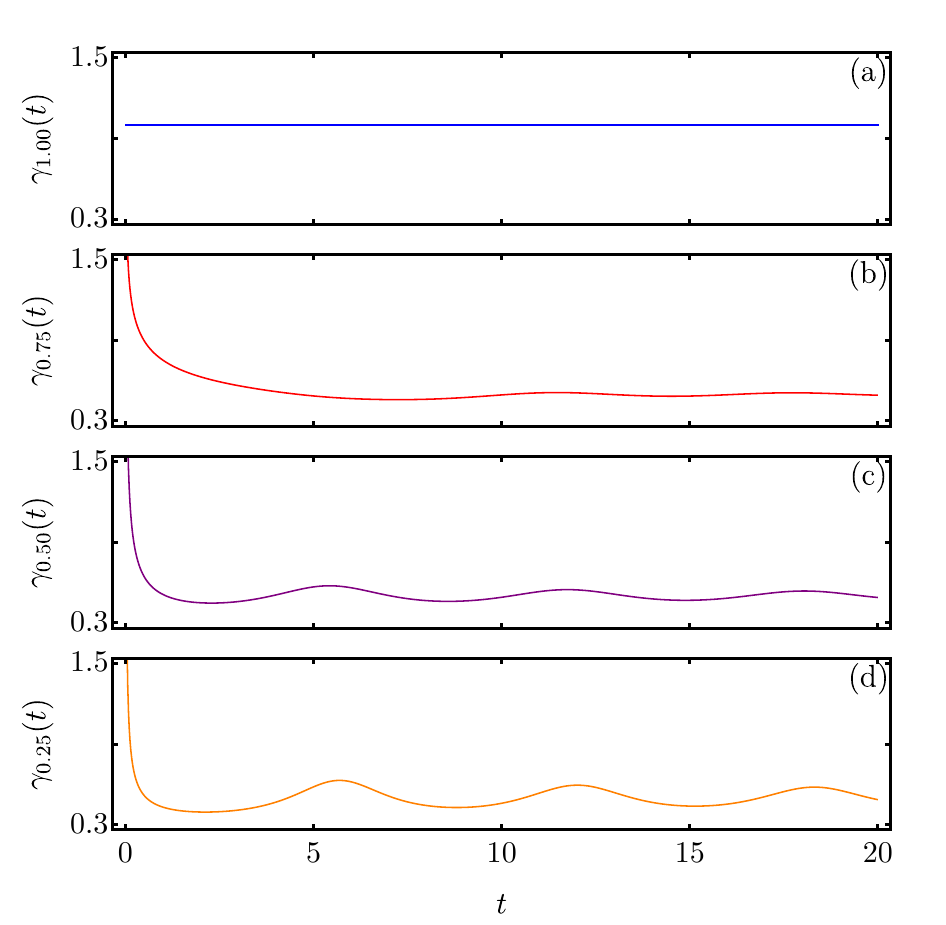}
  \caption{
    The coupling parameter $\gamma_\alpha(t)$, which results in $W_\alpha (t)$ in a non-fractional JC model with the initial state $|e,0\rangle$.
     The ``fractional'' coupling parameter is assumed to be $\mu_\alpha=1$ and
  we consider: (a) $\alpha=1.00$, (b) $\alpha=0.75$, (c) $\alpha=0.50$,  and (d) $\alpha=0.25$.
The effects of fractional-time on the atomic population manifest as a strong initial intensity at the onset of the interaction, followed by a relatively weak coupling whose strength varies subtly over time.
  }
  \label{fig:ipa}
\end{figure}

The coupling profiles obtained through IPA, Eq. \eqref{eq:ipa_formula},
are presented in Fig. \ref{fig:ipa}, for different values of $\alpha$.
As expected, when $\alpha=1.00$, the coupling is
constant, $\gamma_{1.00}(t)=\mu_{1.00}$, as no modulation is needed.
However, for $\alpha < 1.00$, the coupling strength presents a characteristic behavior.
Initially, the effects of the fractional parameter $\alpha$ on the atomic population manifest themselves as a strong coupling intensity that diverges at $t=0$ and corresponds to the powerful initial atom-field interaction responsible for the transient effects.
Subsequently, the coupling settles into a much weaker and slower oscillating value $\gamma_\alpha(t)\approx 0.5$, which accounts for the elongated periods of the RO. Furthermore, by comparing Figs. \ref{fig:ipa}(b), (c), and (d), we observe that as the fractional order $\alpha$ decreases, this initial strength becomes sharper and the amplitude of the subsequent oscillations in $\gamma_\alpha(t)$ increases.

The IPA provides a clear interpretation for the atomic dynamics originating from a Fock state.
To develop a complementary understanding of the field's evolution, we now proceed to an analysis based on an initial coherent state.
This scenario is suited for examining the rich structure of the photon statistics, which provides a direct probe of the cavity field's quantum state.

\section{Photon statistics}
\label{sec:photon}
This section investigates the photon statistics of the FTJC model when the cavity is prepared in an initial coherent state.  In contrast to the previous analysis, we now set the initial state to
\begin{equation}\label{coh_st}
    |\psi_\alpha(0)\rangle=|e,\beta \rangle = e^{-\frac{|\beta|^2}{2}} \sum_{n=0}^\infty \frac{\beta^n}{\sqrt {n!}} |e,n\rangle.
\end{equation}
Coherent states of light are a cornerstone in quantum optics, as they represent a minimum-uncertainty state \cite{Glauber1963c}.
Characterized by a Poissonian distribution, they are more readily accessible in experiments \cite{GERRY2005,Zhang2024,Brune1996}.
In the standard JC model, they ultimately lead to the collapse and revival of the RO.

For the following simulations, we set the coherent state amplitude to $\beta=3$, corresponding to the initial average photon number of $\langle \hat{n}_\alpha (0) \rangle=9$.
Although we have previously considered fractional orders $\alpha=1.00$, $\alpha=0.75$, $\alpha=0.50$, and $\alpha=0.25$, the time evolution for $\alpha=0.25$ exhibits significant rapid oscillations when the initial state is coherent. To mitigate this, we instead use $\alpha = 0.40$ in the present section.

\subsection{Average photon number and parity}
\label{sec:average}

\begin{figure}[t]
  \centering
  \includegraphics[width=1\linewidth]{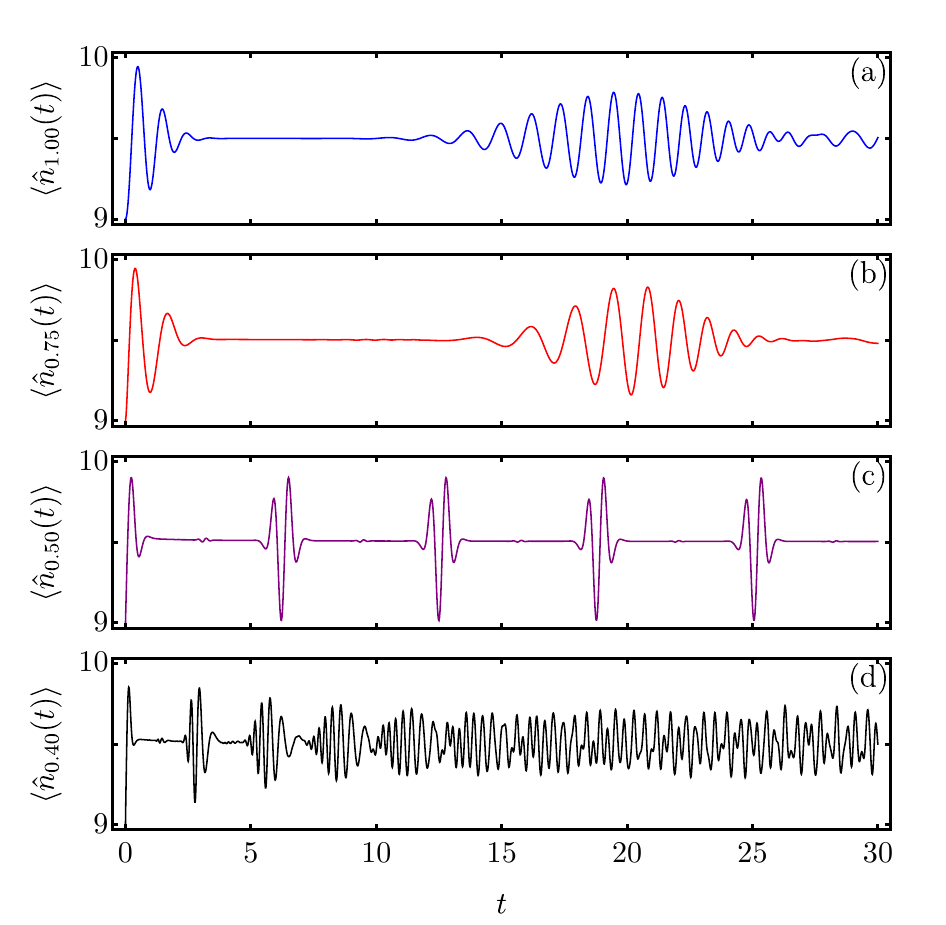}
  \caption{
    The average photon number for the FTJC model, considering the initial state $|\psi_{\alpha}(0)\rangle=|e,\beta\rangle$, fixing $\beta=3$ and $\mu_\alpha=1$, for different values of $\alpha$:
    (a) $\alpha = 1.00$, (b) $\alpha = 0.75$, (c) $\alpha = 0.50$,  and (d) $\alpha = 0.40$.
    The initial values of the Dyson map used are $\kappa_\alpha(0)=\lambda_\alpha(0)=0$ and $\Lambda_\alpha(0)=1$.
    As shown in the plots, different values of $\alpha$ yield distinct characteristics in the quantities considered.
  }
  \label{fig:average_n}
\end{figure}

\begin{figure}[t]
  \centering
  \includegraphics[width=1\linewidth]{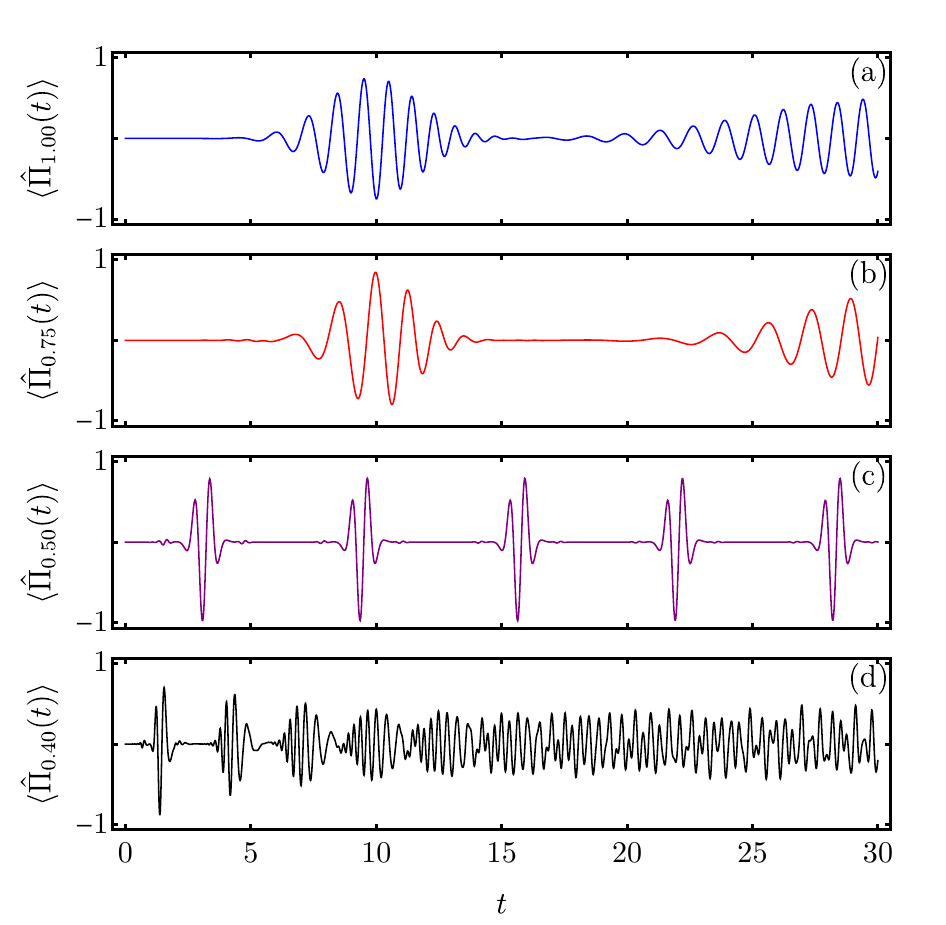}
  \caption{
    The expected value of the parity operator for the FTJC model,
    considering the initial state $|\psi_{\alpha}(0)\rangle=|e,\beta\rangle$, fixing $\beta=3$ and $\mu_\alpha=1$, for different values of $\alpha$:
  (a) $\alpha = 1.00$, (b) $\alpha = 0.75$, (c) $\alpha = 0.50$, and (d) $\alpha = 0.40$.
    The initial values of the Dyson map used are $\kappa_\alpha(0)=\lambda_\alpha(0)=0$ and $\Lambda_\alpha(0)=1$.
    In the quiescent moments of  $\langle \hat{n}_{\alpha}(t)\rangle$, even-odd shifting occurs.
  }
  \label{fig:parity}
\end{figure}

To characterize the state of the cavity field, we first analyze two key quantities: the average photon number and the field parity.
The average photon number
$\langle \hat{n}_\alpha (t)\rangle$, with
$\hat{n}_\alpha = \hat{a}^\dagger \hat{a}$, is a primary indicator of
the field's energy and intensity \cite{Phoenix1988}, with its initial
value defined by the coherent state amplitude $\langle \hat{n}_\alpha
(0) \rangle = |\beta|^2$. 
To further probe the quantum nature of the field, we examine the parity operator $\hat{\Pi}_\alpha=(-1)^{\hat{n}_\alpha}$, which measures the evenness or oddness of the photon number distribution \cite{GERRY2005, Birrittella2015} and serves as an indicator of non-classicality in the cavity mode \cite{gerry2010parity}.
The expectation values of these operators are calculated from the state amplitudes as follows
\begin{subequations}
\begin{align}
    \langle \hat{n}_\alpha (t)\rangle &=\sum_{n=0}^\infty \left[n |A_{e,n}^\alpha (t)|^2 + (n+1) |A_{g,n}^\alpha (t)|^2 \right],
    \\
    \langle \hat{\Pi}_\alpha (t)\rangle &=\sum_{n=0}^\infty (-1)^n \left[ |A_{e,n}^\alpha (t)|^2 - |A_{g,n}^\alpha (t)|^2 \right].
\end{align}
\end{subequations}
The dynamics of these two quantities are presented in Fig. \ref{fig:average_n} and Fig. \ref{fig:parity}, respectively, for the coupling strength $\mu_{\alpha}=1$. For the standard case ($\alpha=1.00$), the average photon number exhibits the well-known collapse and revival structure, Fig. \ref{fig:average_n}(a), thoroughly explained in the literature \cite{Scully1997,GERRY2005}.
The parity operator reveals complementary dynamics: during the quiescent collapse region of $\langle \hat{n}_{1.00}(t)\rangle$, the parity oscillates rapidly, Fig. \ref{fig:parity}(a), a feature ascribed to interference effects in the field's evolution \cite{Birrittella2015}.
As plotted in Figs. \ref{fig:average_n}(b) and \ref{fig:parity}(b), the dynamics for $\alpha=0.75$ are qualitatively similar to the standard case, but the evolution is slower, resulting in a more prolonged collapse and a delayed, less pronounced revival.

On the other hand, we notice a completely different behavior when
$\alpha=0.50$. The mean photon number displays prominent oscillations at
even multiples of $\pi$ ($t=q \pi$, $q=0,2,4\dots$), interspersed with
quiescent intervals where it remains close to an intermediate value, as
shown in Fig. \ref{fig:average_n}(c).
It is precisely during these quiescent periods, at odd multiples of $\pi$ ($t=m \pi$, $m=1,3,5,\dots$), that the parity oscillates, indicating periodic shifts in the photon number distribution, observed in Fig. \ref{fig:parity}(c). For fractional orders below this, such as $\alpha=0.40$, the evolution of both the mean photon number and the parity becomes highly irregular, with the aperiodic behavior of the latter being a direct consequence of the former.
We add that, when $\alpha=0.25$, periodicity is again obtained, however, accompanied by fast RO.

\subsection{Mandel parameter}
The $Q$-Mandel parameter \cite{Mandel1982} is a photon statistics measure,
employed to study the type of distribution associated with a given state of light,
depending on the expected value and uncertainty, denoted by $\Delta ^2 \hat{n}_\alpha(t)$, of the number operator. Governed by
\begin{equation} \label{eq:mandel}
    Q_\alpha (t) = \frac{\Delta ^2 \hat{n}_\alpha(t)}{\langle \hat{n}_\alpha (t) \rangle} - 1,
\end{equation}
the parameter classifies the light source as Poissonian $(Q_\alpha (t)=0)$, super-Poissonian $(Q_\alpha (t)>0)$ and sub-Poissonian $(Q_\alpha (t)<0)$.
The latter is a signal of non-classicality \cite{larson2021},
while the former two are associated with classical light sources, such as coherent or thermal ones.

\begin{figure}[t]
  \centering
  \includegraphics[width=1\linewidth]{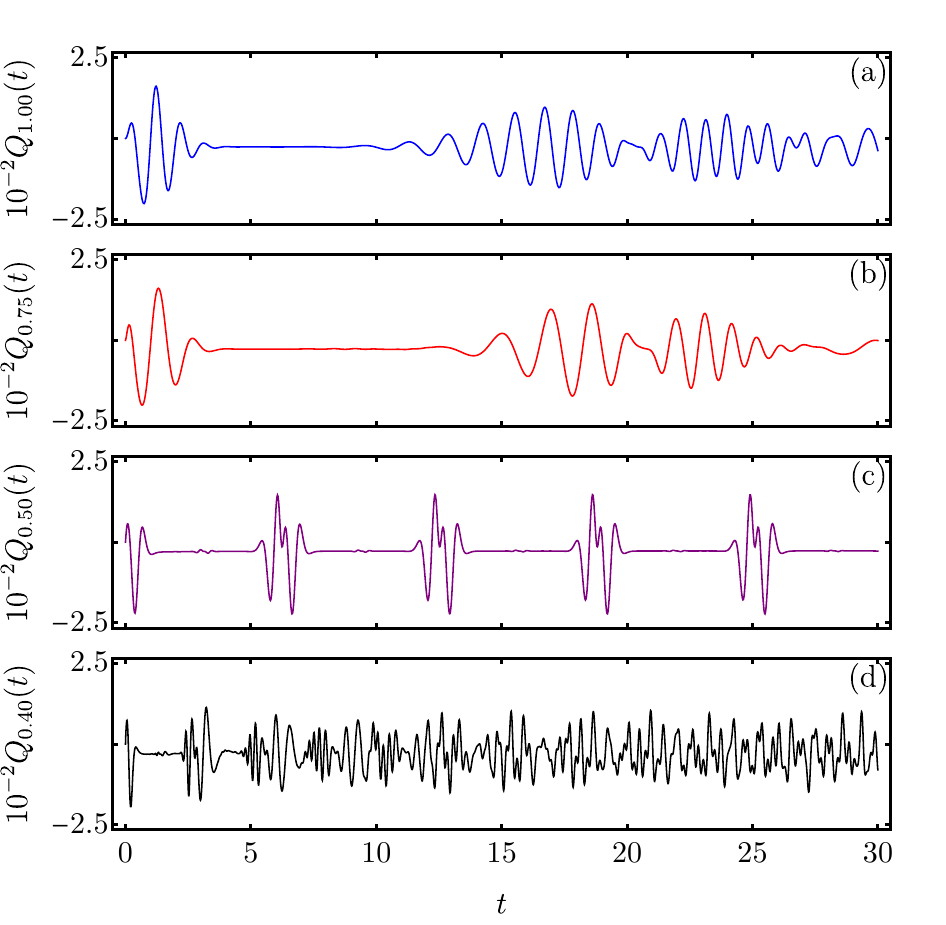}
  \caption{
    The $Q$-Mandel parameter $Q_\alpha(t)$ for the FTJC model, considering the initial state $|\psi_{\alpha}(0)\rangle=|e,\beta\rangle$, fixing $\beta=3$ and $\mu_\alpha=1$, for different values of $\alpha$:
  (a) $\alpha = 1.00$, (b) $\alpha = 0.75$, (c) $\alpha = 0.50$, and (d) $\alpha = 0.40$.
    The initial values of the Dyson map used are $\kappa_\alpha(0)=\lambda_\alpha(0)=0$ and $\Lambda_\alpha(0)=1$.
  }
  \label{fig:mandel}
\end{figure}

In Fig. \ref{fig:mandel}, we display the Mandel $Q$-parameter for the FTJC model.
For $\alpha = 1.00$, a pattern similar to the collapses and revivals emerges, modulating the Mandel $Q$-parameter between super-Poissonian and sub-Poissonian regimes in the initial oscillations and revival.
However, during the collapse, the dynamics exhibits a slight predominance of sub-Poissonian behavior, as indicated by $Q_{1.00}(t) < 0$.
When $\alpha=0.75$, we observe somewhat analogous behavior, but with fewer oscillations.
However, for $\alpha=0.50$, the periodic behavior is again evident, with the system stabilizing in a sub-Poissonian state, $Q_{0.50}(t)<0$, during the quiescent intervals, with oscillations occurring when the respective expectation value of the number operator varies.
A key finding is that the Mandel parameter for $\alpha=0.50$ reaches more negative values than in the standard case, indicating that this fractional order generates a field with stronger non-classical character.
When $\alpha = 0.40$, we observe aperiodic oscillations around a negative value, but with a smaller amplitude compared to the $\alpha = 0.50$ case.

\subsection{Squeezing}
\label{sec:squeezing}

Next, we analyze the squeezing properties of the field quadrature operators \cite{Scully1997}
\begin{equation}
    \begin{aligned}
        \hat{X}_\alpha&=\frac{1}{2}(\hat{a}+\hat{a}^{\dagger}), \\
        \hat{Y}_\alpha&=\frac{1}{2i}(\hat{a}-\hat{a}^{\dagger}).
    \end{aligned}
\end{equation}
These are conjugate operators that satisfy the uncertainty relation:
\begin{equation}
     \Delta^{2}\hat{X}_\alpha\Delta^{2}\hat{Y}_\alpha \geqslant\frac{1}{16}.
\end{equation}
Squeezing is a signature of non-classicality and occurs when the variance of one quadrature falls below the limit of $1/4$, i.e.
\begin{equation}
      \Delta^{2}\hat{X}_\alpha<\frac{1}{4} \mbox{ or } \Delta^{2}\hat{Y}_\alpha<\frac{1}{4}.
\end{equation}
We are interested in the time evolution of the variance $\Delta^{2}\hat{X}_\alpha$,
and how it is affected by the presence of the fractional order time derivative.
Squeezing is known to appear in the standard JC model scenario \cite{Meystre1982,Knight1986,Kuklinski1988},
as a consequence of the nonlinear terms in the time evolution operator \cite{Loudon1987}.
The variance  $\Delta^{2}\hat{X}_\alpha$ can be written as
\begin{equation}
\begin{aligned}
\Delta^{2}\hat{X}_\alpha (t) = \frac{1}{2} \Biggl\{\Re\left[\Delta^{2}\hat{a}(t)\right] + \langle \hat{n}_\alpha (t)\rangle - |\langle\hat{a}(t)\rangle|^{2} +\frac{1}{2} \Biggl\}.
\end{aligned}
\end{equation}

\begin{figure}[t]
  \centering
  \includegraphics[width=1\linewidth]{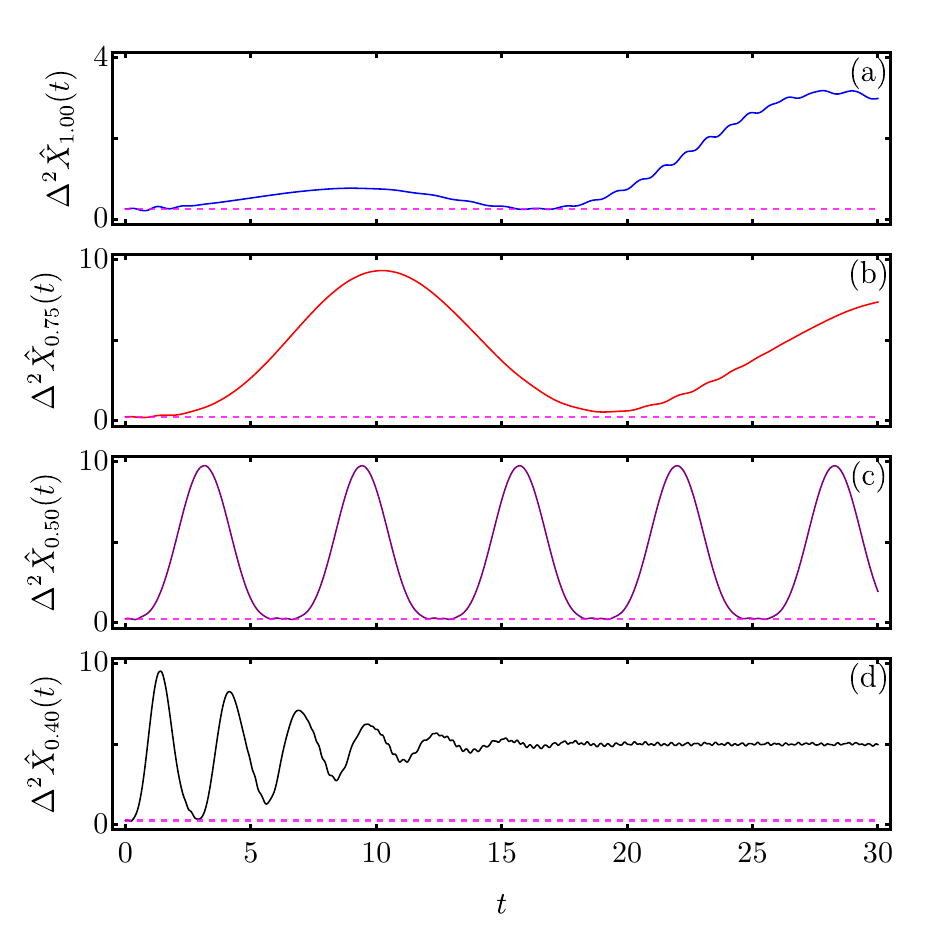}
  \caption{
    The variance of the quadrature operator for the FTJC model, considering the initial state $|\psi_{\alpha}(0)\rangle=|e,\beta\rangle$, fixing $\beta=3$ and $\mu_\alpha=1$, for different values of $\alpha$:
  (a) $\alpha = 1.00$, (b) $\alpha = 0.75$, (c) $\alpha = 0.50$, and  (d) $\alpha = 0.40$.
    The initial values of the Dyson map used are $\kappa_\alpha(0)=\lambda_\alpha(0)=0$ and $\Lambda_\alpha(0)=1$.
    The dashed magenta line represents the value $\Delta^2\hat{X}_\alpha(t)=1/4$.
    We observe periodic squeezing for $\alpha=0.50$ and an asymptotic value in the variance when $\alpha=0.40$.
  }
  \label{fig:var}
\end{figure}

\begin{figure}[t!]
  \centering
  \includegraphics[width=1\linewidth]{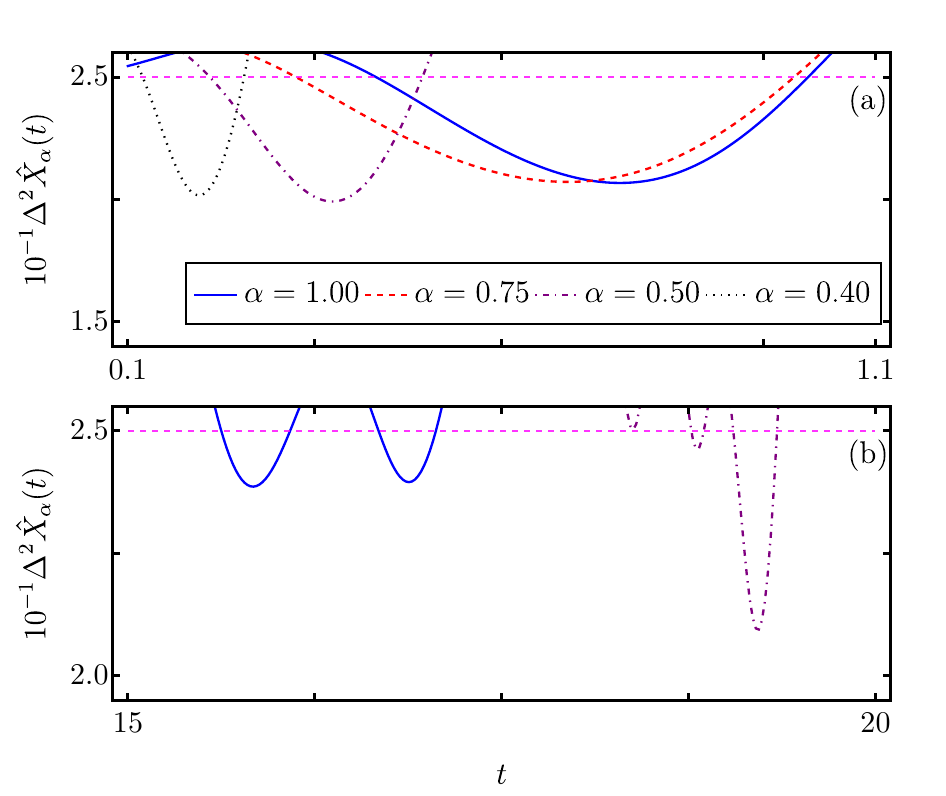}
  \caption{
    The variance of the quadrature operator for the FTJC model, considering the initial state $|\psi_{\alpha}(0)\rangle=|e,\beta\rangle$, fixing $\beta=3$ and $\mu_\alpha=1$, for specific time intervals:
  (a) $0.1<t<1.1$, and (b) $15<t<20$.
    We considered different values of $\alpha$:  $\alpha=1.00$ (solid blue line), $\alpha=0.75$ (dashed red line), $\alpha=0.50$ (dot-dashed purple  line), and $\alpha=0.40$ (dotted black line).
    The initial values of the Dyson map used are $\kappa_\alpha(0)=\lambda_\alpha(0)=0$ and $\Lambda_\alpha(0)=1$.
  }
  \label{fig:zoom}
\end{figure}

 In Fig. \ref{fig:var}, we present the variance in the fractional-time scenario, for different values of $\alpha$.
 For $\alpha = 1.00$, we observe squeezing during the initial oscillations, followed by an increase in $ \Delta^{2}\hat{X}_{1.00} (t)$.
 Subsequently, in the interval $10 < t < 20$ and with the onset of revival in Fig.\ref{fig:average_n}(a), we again perceive non-classical values for the variance \cite{Meystre1982,Loudon1987}.
This behavior is linked to the number-phase uncertainty relationship: intervals of large photon number uncertainty (super-Poissonian statistics, $Q_\alpha(t)>0$) correlate with intervals where the phase-sensitive quadrature variance can be squeezed \cite{Schleich2001}.
When $\alpha = 0.75$, squeezing is present only at the initial times, and it is noticeable that the variance reaches higher values between the two minima in this case, as a consequence of the slower start to the revival.
On the other hand, for $\alpha = 0.50$, squeezing appears periodically at $t=q \pi$, when oscillations occur in Figs. \ref{fig:average_n}(c) and \ref{fig:mandel}(c).
Conversely, at the instants $t=m \pi$, the variance reaches high values.
For $\alpha = 0.40$, the variance settles to an asymptotic value after an initial period of squeezing.
This loss of strong squeezing is consistent with the $Q$-Mandel parameter approaching zero, indicating that photon statistics are becoming more classical (Poissonian-like).
Fig. \ref{fig:zoom} presents a magnified view of the same quantity, highlighting its behavior during the collapse (a) and revival (b) stages.
Our results indicate that squeezing occurs earlier for $\alpha = 0.40$ compared to the standard case ($\alpha = 1.00$).
The highest intensity of squeezing is achieved for $\alpha = 0.50$, whereas the least pronounced squeezing occurs at $\alpha = 0.75$.
This indicates that different fractional derivative orders provide access to different regimes, classical and non-classical, regarding quadrature squeezing.


\subsection{Husimi function}
\label{sec:husimi}

Phase-space quasi-probability distributions provide an intuitive representation of the quantum state of the electromagnetic field.
In this work, we consider the Husimi function \cite{Husimi1940,Kano1965,Cahill1969}, a positive-definite phase-space distribution \cite{Cartwright1976}, obtained by projecting the quantum state of the field onto the overcomplete basis of coherent states.
This function has been computed for the lossless \cite{Eiselt1989,Buzek1992} and damped JC model \cite{Eiselt1989,Eiselt1991}.
In addition, it has been used for the two-photon \cite{Buzek1993}, driven field \cite{Alsing1991}, and time-dependent \cite{Fang1998} extensions of the standard JC model.

The Husimi function $\mathcal{Q}_{\alpha}(t)$ is defined as the expectation value of the reduced density matrix of the field $\hat{\rho}_\alpha^{F}(t)$, with respect to an arbitrary coherent state $|\gamma\rangle$
\begin{equation}
\mathcal{Q}_\alpha(t)=\frac{1}{\pi}\langle\gamma|\hat{\rho}_\alpha^{F}(t)|\gamma\rangle,
\end{equation}
where $\hat{\rho}_\alpha^{F}(t)$ is obtained through the partial trace over the atomic degrees of freedom,
$\hat{\rho}^{F}_\alpha(t) = \Tr_A{\left[\hat{\rho}_\alpha(t)\right]}$.
The Husimi function obeys the normalization condition $\int\mathcal{Q}_\alpha(t)d^{2}\gamma = 1$.

\begin{figure}[t]
  \centering
  \includegraphics[width=1\linewidth]{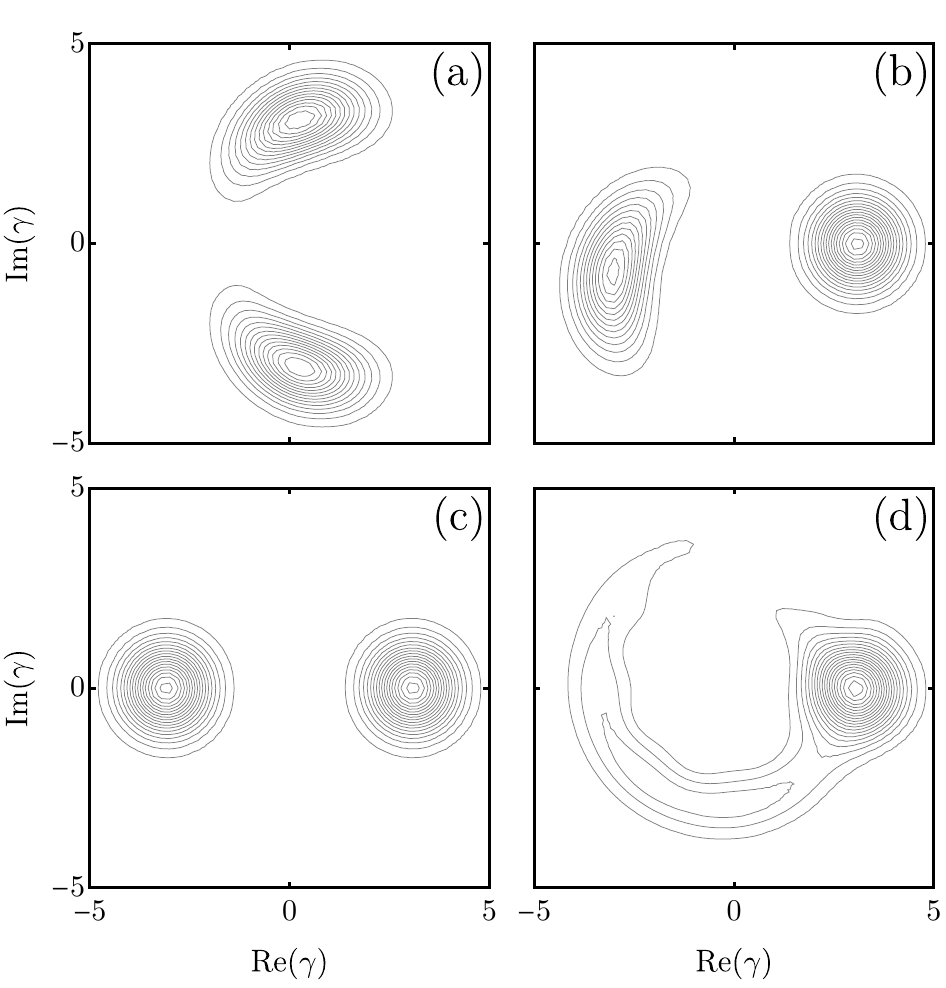}
  \caption{
    The Husimi function $\mathcal{Q}_\alpha(t)$ for the FTJC model, considering the initial state $|\psi_{\alpha}(0)\rangle=|e,\beta\rangle$, fixing $\beta=3$, $\mu_\alpha=1$ and $t=t_r/2=\pi |\beta|/\mu_\alpha$, for different values of $\alpha$:
  (a) $\alpha = 1.00$, (b) $\alpha = 0.75$, (c) $\alpha = 0.50$,  and (d) $\alpha = 0.40$.
    The initial values of the Dyson map used are $\kappa_\alpha(0)=\lambda_\alpha(0)=0$ and $\Lambda_\alpha(0)=1$.
    Particularly interesting is the formation of two Gaussian-like distributions when $\alpha=0.50$.
  }
  \label{fig:husimi}
\end{figure}

In Fig. \ref{fig:husimi}, we present the Husimi function for different values of $\alpha$ at $t=t_r/2=|\beta| \pi/\mu_\alpha$.
In the standard JC model ($\alpha=1$), $t_r$ is the first revival time \cite{larson2021}, and the halfway point, $t_r/2$, is significant as it is where the field evolves into a macroscopic quantum superposition known as a Schrödinger cat state \cite{Phoenix1991}.
As seen in Fig. \ref{fig:husimi}(a), the Husimi function evolves by bifurcating into two distinct blobs that rotate in opposite directions along a circle of radius $|\beta|^2$, centered at the origin \cite{SHORE1993}.
At the given instant, the two blobs present the same intensity but different phases,
characterizing the aforementioned Schr\"odinger cat state.

On the other hand, when $\alpha < 1.00$, the phase-space dynamics are distinct.
The initial distribution splits into two blobs; one remains stationary at the initial position, displaying a Gaussian aspect,
while the other traverses the circle of radius $|\beta|^2$.
For $\alpha=0.75$, the distributions are asymmetric, while for $\alpha=0.50$
we recover the cat state, now characterized by two Gaussian distributions.
Furthermore, a connection between the field's phase-space evolution and the periodic even-odd transitions can be observed in Fig. \ref{fig:husimi}(c).
The formation of these cat states is periodic, coinciding with the parity oscillations at
$t = m\pi$, as shown in Fig. \ref{fig:parity}(c).
In our simulations, this periodicity persisted throughout the entire time interval studied.
For $\alpha = 0.40$, the two parts of the Husimi function have joined at the considered instant, although a peak remains at the initial point of the plane.
Furthermore, a similar merging occurs at a later time for $\alpha = 0.75$.

Although no direct equivalence between the underlying dynamics can be claimed, the phase-space distribution for $\alpha=0.50$ is remarkably similar to that occurring in an entirely different system: the two-photon JC model with a Stark shift \cite{Puri1988,Buzek1993}.
In both cases, the field evolves into a Schr\"{o}dinger cat state composed of two distinct Gaussian components.
This correspondence, while not evidence of a shared physical origin, serves as a compelling qualitative benchmark for the highly non-trivial effects of the fractional-time evolution.

\section{Conclusions}
\label{sec:Conc}
In this work, we investigate the statistical properties of the fractional-time Jaynes-Cummings model by employing the unitary framework proposed in  Ref. \cite{Cius2024}.
Our analysis first addressed an initial Fock state, revealing that the evolution manifests itself as transient effects in the atomic population and atom-field entanglement, followed by Rabi oscillations at a reduced frequency compared to the standard model.
In addition, a higher sensitivity to the coupling parameter is observed when $\alpha<1.00$.

To replicate fractional-time aspects of the dynamics, we employed the Inverse Problem Approach.
This method yielded an effective time-dependent coupling for a standard Jaynes-Cummings model, characterized by a divergent coupling at $t=0$, which accounts for transient effects and a subsequent weaker interaction corresponding to the reduced oscillation frequency.
The investigation was then extended to an initial coherent state of the cavity, revealing distinct dynamical regimes in the photon statistics governed by the fractional order $\alpha$.
For example, when $\alpha=0.75$, a behavior somewhat similar to the standard framework is observed in the expectation value of the number and parity operators, as well as in the  $Q$-Mandel parameter, with slower oscillations.
Remarkably, the squeezing phenomenon is diminished.
On the other hand, periodicity is observed in the different quantities analyzed when $\alpha=0.50$.
In fact, this leads to an enhancement of non-classical features, including stronger sub-Poissonian statistics, more pronounced squeezing, and the recurrent formation of a Schrödinger cat state, composed of two distinct Gaussian distributions in phase-space.
For $\alpha=0.40$, an aperiodic behavior emerges, resulting in an asymptotic value in the variance of the quadrature operator.
These findings not only deepen the understanding of the fractional Jaynes-Cummings model but also open avenues for exploring the physical significance and potential applications of fractional calculus in other quantum systems.

\section*{Acknowledgements}
The authors thank Alison A. Silva and Enrique C. Gabrick for their
valuable comments and clarifications that significantly improved the
work. 
T.T.T. acknowledges financial support from the Coordenação de
Aperfeiçoamento de Pessoal de N\'{i}vel Superior (CAPES, Finance Code
001).
D.C. would like to acknowledge financial support from Instituto
Serrapilheira and the Pró-Reitoria de Pesquisa e Inovação (PRPI) of the
Universidade de São Paulo (USP) through the Programa de Estímulo à
Supervisão de Pós-Doutorandos por Jovens Pesquisadores.
F.M.A. and A.S.M.C. acknowledge financial support from Fundação Araucária (Project No. 305).
F.M.A. also acknowledges financial support by CNPq Grant No. 313124/2023-0.

\onecolumngrid

\appendix
\section{Explicit Solutions for the Unitary Framework}
\label{app:solutions}
This appendix provides the explicit expressions required to construct the time-dependent Dyson map $\hat{\eta}_{\alpha}^{(n)}(t)$ and the resulting unitary time-evolution operator $\hat{u}^{(n)}_{\alpha}(t)$.
These expressions are found in Refs. \cite{cius22frac,Cius2024}.

The explicit solutions for the time-dependent parameters of the Dyson map in Eq. \eqref{eq:etamat} are given by:
\begin{subequations}
\label{eq:dyson_solutions_app}
\begin{align}
    \kappa_{\alpha}^{(n)}(t) &= \kappa_{\alpha}^{(n)}(0) -\frac{1}{2}\mathrm{Re}[\ln{D_{\alpha}^{(n)}}(t)], \\
    \chi_{\alpha}^{(n)}(t) &= \frac{|\zeta_{\alpha,+}^{(n)}(t)|^{2} + |\zeta_{\alpha,-}^{(n)}(t)|^{2} + \Lambda_{\alpha}^{(n)}(0)e^{\mathrm{Re}[\ln{D_{\alpha}^{(n)}}(t)]}}{ |\xi_{\alpha,+}^{(n)}(t)|^{2} + |\xi_{\alpha,-}^{(n)}(t)|^{2} + \Lambda_{\alpha}^{(n)}(0)e^{\mathrm{Re}[\ln{D_{\alpha}^{(n)}}(t)]}}, \\
    \lambda_{\alpha}^{(n)}(t) &= \frac{-\left[\xi_{\alpha,+}^{(n)}(t)[\zeta_{\alpha,+}^{(n)}(t)]^{\ast} + \xi_{\alpha,-}^{(n)}(t)[\zeta_{\alpha,-}^{(n)}(t)]^{\ast}\right]}{|\xi_{\alpha,+}^{(n)}(t)|^{2} + |\xi_{\alpha,-}^{(n)}(t)|^{2} + \Lambda_{\alpha}^{(n)}(0)e^{\mathrm{Re}[\ln{D_{\alpha}^{(n)}}(t)]}}.
\end{align}
\end{subequations}

The auxiliary functions used in the above expressions are defined as follows. The function $D_{\alpha}^{(n)}(t)$ reads
\begin{equation}
    D_{\alpha}^{(n)}(t) = [\mathcal{C}^{(n)}_{\alpha}(t)]^{2} - (-1)^{-\alpha}[\mathcal{S}^{(n)}_{\alpha}(t)]^{2}.
\end{equation}

The functions $\zeta_{\alpha,\pm}^{(n)}(t)$ and $\xi_{\alpha,\pm}^{(n)}(t)$ contain the time-dependence from the non-unitary evolution and the initial conditions of the Dyson map parameters at $t=0$:
\begin{subequations}
\begin{align}
    \zeta_{\alpha,+}^{(n)}(t) &=  i^{-\alpha}\mathcal{S}_{\alpha}^{(n)}(t) - [\lambda_{\alpha}^{(n)}(0)]^{\ast} \mathcal{C}_{\alpha}^{(n)}(t), \\
    \zeta_{\alpha,-}^{(n)}(t) &= i^{-\alpha}\lambda_{\alpha}^{(n)}(0) \mathcal{S}_{\alpha}^{(n)}(t) - \chi_{\alpha}^{(n)}(0) \mathcal{C}_{\alpha}^{(n)}(t), \\
    \xi_{\alpha,+}^{(n)}(t) &=  \mathcal{C}_{\alpha}^{(n)}(t) - i^{-\alpha}[\lambda_{\alpha}^{(n)}(0)]^{\ast} \mathcal{S}_{\alpha}^{(n)}(t) ,\\
    \xi_{\alpha,-}^{(n)}(t) &= \lambda_{\alpha}^{(n)}(0) \mathcal{C}_{\alpha}^{(n)}(t) -  i^{-\alpha}\chi_{\alpha}^{(n)}(0) \mathcal{S}_{\alpha}^{(n)}(t).
\end{align}
\end{subequations}

Finally, the components of the unitary operator $\hat{u}^{(n)}_{\alpha}(t)$ in Eq.~\eqref{eq:u_matrix_final} are constructed from these functions.
The overall phase $\delta_{\alpha}^{(n)}(t)$ is
\begin{align}
    \delta_{\alpha}^{(n)}(t) = \frac{1}{2}\mathrm{Im}[\ln{D_{\alpha}^{(n)}(t)}],
\end{align}
and the evolution amplitudes $\varpi_{\alpha,\pm}^{(n)}(t)$ are derived from the intermediate function $\nu_{\alpha,\pm}^{(n)}(t)$:
\begin{subequations}
\begin{align}
\varpi_{\alpha,\pm}^{(n)}(t) &= \pm e^{i\delta_{\alpha}^{(n)}(t)}[\nu_{\alpha,\mp}^{(n)}(t)]^{\ast}, \\
\nu_{\alpha,\pm}^{(n)}(t) &= \pm\frac{e^{\kappa_{\alpha}^{(n)}(t)-\kappa_{\alpha}^{(n)}(0)}}{\sqrt{\Lambda_{\alpha}^{(n)}(t)\Lambda_{\alpha}^{(n)}(0)}}
\left[ \zeta_{\alpha,\pm}^{(n)}(t) + \left[\lambda_{\alpha}^{(n)}(t)\right]^{\ast} \xi_{\alpha,\pm}^{(n)}(t) \right].
\end{align}
\end{subequations}

\section{Positive Coupling Strength via the Inverse Problem Approach}
\label{app:ipa}

This appendix demonstrates that the population inversion, $W(t)$, is governed by the magnitude of the time-dependent coupling, $\gamma(t)$, thereby justifying the choice of a strictly positive, real coupling in the main text.
Let us consider the Schrödinger equation for a standard JC model in the interaction picture,
\begin{equation}
i\partial_{t}\vert\psi(t)\rangle = \hat{V}(t)\vert\psi(t)\rangle,
\end{equation}
with the Hamiltonian
\begin{equation}
\hat{V}(t)= \gamma(t)\hat{\sigma}^{+}\hat{a} + \gamma^{\ast}(t)\hat{\sigma}^{-}\hat{a}^{\dagger}.
\end{equation}

For a system initialized in the state $\vert\psi(0)\rangle = \vert e,0\rangle$, the solution is constrained to the single-excitation subspace and has the general form
\begin{equation}
\label{eq:app_state}
\vert\psi(t)\rangle = A_{e}(t)\vert e,0\rangle + A_{g}(t)\vert g,1\rangle.
\end{equation}
The probability amplitudes are related to the population inversion $W(t)$ by
\begin{subequations}
    \begin{align}
        |A_{e}(t)|^{2}+|A_{g}(t)|^{2}&=1,
        \\
        |A_{e}(t)|^{2}-|A_{g}(t)|^{2}&=W(t),
    \end{align}
\end{subequations}
Solving for the magnitudes yields
\begin{subequations}
\label{eq:app_amplitude}
    \begin{align}
        |A_{e}(t)|&=\sqrt{\frac{1+W(t)}{2}},
        \\
        |A_{g}(t)|&=\sqrt{\frac{1-W(t)}{2}},
    \end{align}
\end{subequations}
which requires  $|W(t)|\leqslant 1$.

Substituting the state ansatz, Eq. \eqref{eq:app_state}, into the Schrödinger equation and projecting onto the bare states yields the coupled equations for the amplitudes
\begin{subequations}
    \begin{align}
        i\dot{A}_{e}(t) &= \gamma(t)A_{g}(t),
        \\
        i\dot{A}_{g}(t) &= \gamma^\ast(t)A_{e}(t).
    \end{align}
\end{subequations}
From the first of these equations, we can write
\begin{equation}
    \gamma(t) = i\frac{\dot{A}_{e}(t)}{A_{g}(t)}.
\end{equation}
Representing the amplitudes in polar form, $A_{k}(t)=|A_{k}(t)|e^{i\theta_{k}(t)}$ for $k=e,g$,  we can set the overall phase such that $\theta_{e}(t)=0$ without loss of generality. Denoting the relative phase as $\theta_{g}\equiv \theta(t)$, we obtain
\begin{equation}
\label{eq:app_gamma0}
    \gamma(t) = i\frac{\partial_{t}{|A_{e}(t)|}}{|A_{g}(t)|}e^{-i\theta(t)}.
\end{equation}
Using the relations from Eq. \eqref{eq:app_amplitude}, we find $\partial_{t}{|A_{e}(t)|} = \dot{W}(t)/[4|A_{e}(t)|]$. Substituting this and Eq. \eqref{eq:app_amplitude} into Eq. \eqref{eq:app_gamma0} yields the general form of the complex coupling
\begin{equation}
    \gamma(t) = \frac{\dot{W}(t)}{2\sqrt{1-W^{2}(t)}}e^{-i[\theta(t)-\pi/2]}.
\end{equation}
The sign of $\dot{W}(t)$  can be absorbed into the phase, allowing us to write the coupling in terms of its magnitude and a phase factor
\begin{equation}
    \gamma(t) = \left\vert
    \frac{\dot{W}(t)}{2\sqrt{1-W^{2}(t)}}
    \right\vert
    e^{i\theta_{\gamma}(t)},
\end{equation}
where $\theta_\gamma(t) =\pi/2 + \mathfrak{h}(t)\pi -\theta(t)$, with
\begin{equation}
    \mathfrak{h}(t) =
    \Bigg\{
    \begin{array}{cc}
        1, &  \dot{W}(t)\geqslant 0 \\
        0, & \dot{W}(t)< 0
    \end{array},
\end{equation}
being the step function.

The dynamical behavior of the population inversion, $W(t)$, is insensitive to the phase of the coupling, $\theta_{\gamma}(t)$.
As shown by the evolution of the probability amplitudes, any choice of $\theta_{\gamma}(t)$ can be compensated by a corresponding shift in the phase of the state, $\theta(t)$. This gauge freedom allows us to define a strictly positive,  real coupling by setting $\theta_{\gamma}(t)=0$, which leads to
\begin{equation}
    \theta(t) = \mathfrak{h}(t)\pi + \frac{\pi}{2}.
\end{equation}
This choice is sufficient to reproduce any desired population inversion dynamics without loss of generality.

\twocolumngrid

\bibliographystyle{apsrev4-2}
%


\begin{thebibliography}{92}%
\makeatletter
\providecommand \@ifxundefined [1]{%
 \@ifx{#1\undefined}
}%
\providecommand \@ifnum [1]{%
 \ifnum #1\expandafter \@firstoftwo
 \else \expandafter \@secondoftwo
 \fi
}%
\providecommand \@ifx [1]{%
 \ifx #1\expandafter \@firstoftwo
 \else \expandafter \@secondoftwo
 \fi
}%
\providecommand \natexlab [1]{#1}%
\providecommand \enquote  [1]{``#1''}%
\providecommand \bibnamefont  [1]{#1}%
\providecommand \bibfnamefont [1]{#1}%
\providecommand \citenamefont [1]{#1}%
\providecommand \href@noop [0]{\@secondoftwo}%
\providecommand \href [0]{\begingroup \@sanitize@url \@href}%
\providecommand \@href[1]{\@@startlink{#1}\@@href}%
\providecommand \@@href[1]{\endgroup#1\@@endlink}%
\providecommand \@sanitize@url [0]{\catcode `\\12\catcode `\$12\catcode
  `\&12\catcode `\#12\catcode `\^12\catcode `\_12\catcode `\%12\relax}%
\providecommand \@@startlink[1]{}%
\providecommand \@@endlink[0]{}%
\providecommand \url  [0]{\begingroup\@sanitize@url \@url }%
\providecommand \@url [1]{\endgroup\@href {#1}{\urlprefix }}%
\providecommand \urlprefix  [0]{URL }%
\providecommand \Eprint [0]{\href }%
\providecommand \doibase [0]{https://doi.org/}%
\providecommand \selectlanguage [0]{\@gobble}%
\providecommand \bibinfo  [0]{\@secondoftwo}%
\providecommand \bibfield  [0]{\@secondoftwo}%
\providecommand \translation [1]{[#1]}%
\providecommand \BibitemOpen [0]{}%
\providecommand \bibitemStop [0]{}%
\providecommand \bibitemNoStop [0]{.\EOS\space}%
\providecommand \EOS [0]{\spacefactor3000\relax}%
\providecommand \BibitemShut  [1]{\csname bibitem#1\endcsname}%
\let\auto@bib@innerbib\@empty
\bibitem [{\citenamefont {Podlubny}(1998)}]{podlubny1998fractional}%
  \BibitemOpen
  \bibfield  {author} {\bibinfo {author} {\bibfnamefont {I.}~\bibnamefont
  {Podlubny}},\ }\href@noop {} {\emph {\bibinfo {title} {Fractional
  differential equations}}}\ (\bibinfo  {publisher} {Academic Press},\ \bibinfo
  {year} {1998})\BibitemShut {NoStop}%
\bibitem [{\citenamefont {Metzler}\ and\ \citenamefont
  {Klafter}(2000)}]{metzler2000random}%
  \BibitemOpen
  \bibfield  {author} {\bibinfo {author} {\bibfnamefont {R.}~\bibnamefont
  {Metzler}}\ and\ \bibinfo {author} {\bibfnamefont {J.}~\bibnamefont
  {Klafter}},\ }\href
  {https://doi.org/https://doi.org/10.1016/S0370-1573(00)00070-3} {\bibfield
  {journal} {\bibinfo  {journal} {Phys. Rep.}\ }\textbf {\bibinfo {volume}
  {339}},\ \bibinfo {pages} {1} (\bibinfo {year} {2000})}\BibitemShut {NoStop}%
\bibitem [{\citenamefont {Zhang}\ \emph {et~al.}(2015)\citenamefont {Zhang},
  \citenamefont {Liu}, \citenamefont {Beli\ifmmode~\acute{c}\else \'{c}\fi{}},
  \citenamefont {Zhong}, \citenamefont {Zhang},\ and\ \citenamefont
  {Xiao}}]{zhang2015}%
  \BibitemOpen
  \bibfield  {author} {\bibinfo {author} {\bibfnamefont {Y.}~\bibnamefont
  {Zhang}}, \bibinfo {author} {\bibfnamefont {X.}~\bibnamefont {Liu}}, \bibinfo
  {author} {\bibfnamefont {M.~R.}\ \bibnamefont {Beli\ifmmode~\acute{c}\else
  \'{c}\fi{}}}, \bibinfo {author} {\bibfnamefont {W.}~\bibnamefont {Zhong}},
  \bibinfo {author} {\bibfnamefont {Y.}~\bibnamefont {Zhang}},\ and\ \bibinfo
  {author} {\bibfnamefont {M.}~\bibnamefont {Xiao}},\ }\href
  {https://doi.org/10.1103/PhysRevLett.115.180403} {\bibfield  {journal}
  {\bibinfo  {journal} {Phys. Rev. Lett.}\ }\textbf {\bibinfo {volume} {115}},\
  \bibinfo {pages} {180403} (\bibinfo {year} {2015})}\BibitemShut {NoStop}%
\bibitem [{\citenamefont {Longhi}(2015)}]{longhi2015fractional}%
  \BibitemOpen
  \bibfield  {author} {\bibinfo {author} {\bibfnamefont {S.}~\bibnamefont
  {Longhi}},\ }\href {https://doi.org/https://doi.org/10.1364/OL.40.001117}
  {\bibfield  {journal} {\bibinfo  {journal} {Opt. Lett}\ }\textbf {\bibinfo
  {volume} {40}},\ \bibinfo {pages} {1117} (\bibinfo {year}
  {2015})}\BibitemShut {NoStop}%
\bibitem [{\citenamefont {Zhang}\ \emph {et~al.}(2016)\citenamefont {Zhang},
  \citenamefont {Zhong}, \citenamefont {Beli{\'c}}, \citenamefont {Zhu},
  \citenamefont {Zhong}, \citenamefont {Zhang}, \citenamefont
  {Christodoulides},\ and\ \citenamefont {Xiao}}]{zhang2016pt}%
  \BibitemOpen
  \bibfield  {author} {\bibinfo {author} {\bibfnamefont {Y.}~\bibnamefont
  {Zhang}}, \bibinfo {author} {\bibfnamefont {H.}~\bibnamefont {Zhong}},
  \bibinfo {author} {\bibfnamefont {M.~R.}\ \bibnamefont {Beli{\'c}}}, \bibinfo
  {author} {\bibfnamefont {Y.}~\bibnamefont {Zhu}}, \bibinfo {author}
  {\bibfnamefont {W.}~\bibnamefont {Zhong}}, \bibinfo {author} {\bibfnamefont
  {Y.}~\bibnamefont {Zhang}}, \bibinfo {author} {\bibfnamefont {D.~N.}\
  \bibnamefont {Christodoulides}},\ and\ \bibinfo {author} {\bibfnamefont
  {M.}~\bibnamefont {Xiao}},\ }\href
  {https://doi.org/https://doi.org/10.1002/lpor.201600037} {\bibfield
  {journal} {\bibinfo  {journal} {Laser Photonics Rev.}\ }\textbf {\bibinfo
  {volume} {10}},\ \bibinfo {pages} {526} (\bibinfo {year} {2016})}\BibitemShut
  {NoStop}%
\bibitem [{\citenamefont {Huang}\ and\ \citenamefont
  {Dong}(2017)}]{huang2017beam}%
  \BibitemOpen
  \bibfield  {author} {\bibinfo {author} {\bibfnamefont {C.}~\bibnamefont
  {Huang}}\ and\ \bibinfo {author} {\bibfnamefont {L.}~\bibnamefont {Dong}},\
  }\href {https://doi.org/https://doi.org/10.1038/s41598-017-05926-5}
  {\bibfield  {journal} {\bibinfo  {journal} {Sci. Rep.}\ }\textbf {\bibinfo
  {volume} {7}},\ \bibinfo {pages} {1} (\bibinfo {year} {2017})}\BibitemShut
  {NoStop}%
\bibitem [{\citenamefont {Colas}(2020)}]{colas2020}%
  \BibitemOpen
  \bibfield  {author} {\bibinfo {author} {\bibfnamefont {D.}~\bibnamefont
  {Colas}},\ }\href {https://doi.org/10.1103/PhysRevResearch.2.033274}
  {\bibfield  {journal} {\bibinfo  {journal} {Phys. Rev. Res.}\ }\textbf
  {\bibinfo {volume} {2}},\ \bibinfo {pages} {033274} (\bibinfo {year}
  {2020})}\BibitemShut {NoStop}%
\bibitem [{\citenamefont {Wu}\ \emph {et~al.}(2010)\citenamefont {Wu},
  \citenamefont {Huang}, \citenamefont {Cheng},\ and\ \citenamefont
  {Hsieh}}]{wu2010}%
  \BibitemOpen
  \bibfield  {author} {\bibinfo {author} {\bibfnamefont {J.-N.}\ \bibnamefont
  {Wu}}, \bibinfo {author} {\bibfnamefont {C.-H.}\ \bibnamefont {Huang}},
  \bibinfo {author} {\bibfnamefont {S.-C.}\ \bibnamefont {Cheng}},\ and\
  \bibinfo {author} {\bibfnamefont {W.-F.}\ \bibnamefont {Hsieh}},\ }\href
  {https://doi.org/10.1103/PhysRevA.81.023827} {\bibfield  {journal} {\bibinfo
  {journal} {Phys. Rev. A}\ }\textbf {\bibinfo {volume} {81}},\ \bibinfo
  {pages} {023827} (\bibinfo {year} {2010})}\BibitemShut {NoStop}%
\bibitem [{\citenamefont {Stickler}(2013)}]{stickler2013}%
  \BibitemOpen
  \bibfield  {author} {\bibinfo {author} {\bibfnamefont {B.~A.}\ \bibnamefont
  {Stickler}},\ }\href {https://doi.org/10.1103/PhysRevE.88.012120} {\bibfield
  {journal} {\bibinfo  {journal} {Phys. Rev. E}\ }\textbf {\bibinfo {volume}
  {88}},\ \bibinfo {pages} {012120} (\bibinfo {year} {2013})}\BibitemShut
  {NoStop}%
\bibitem [{\citenamefont {Stephanovich}\ \emph {et~al.}(2022)\citenamefont
  {Stephanovich}, \citenamefont {Kirichenko}, \citenamefont {Dugaev},
  \citenamefont {Sauco},\ and\ \citenamefont {Brito}}]{stephanovich2022}%
  \BibitemOpen
  \bibfield  {author} {\bibinfo {author} {\bibfnamefont {V.}~\bibnamefont
  {Stephanovich}}, \bibinfo {author} {\bibfnamefont {E.}~\bibnamefont
  {Kirichenko}}, \bibinfo {author} {\bibfnamefont {V.}~\bibnamefont {Dugaev}},
  \bibinfo {author} {\bibfnamefont {J.~H.}\ \bibnamefont {Sauco}},\ and\
  \bibinfo {author} {\bibfnamefont {B.~L.}\ \bibnamefont {Brito}},\ }\href
  {https://doi.org/https://doi.org/10.1038/s41598-022-16597-2} {\bibfield
  {journal} {\bibinfo  {journal} {Sci. Rep.}\ }\textbf {\bibinfo {volume}
  {12}},\ \bibinfo {pages} {12540} (\bibinfo {year} {2022})}\BibitemShut
  {NoStop}%
\bibitem [{\citenamefont {Zhong}\ \emph {et~al.}(2024)\citenamefont {Zhong},
  \citenamefont {Malomed},\ and\ \citenamefont {Yan}}]{zhong2024}%
  \BibitemOpen
  \bibfield  {author} {\bibinfo {author} {\bibfnamefont {M.}~\bibnamefont
  {Zhong}}, \bibinfo {author} {\bibfnamefont {B.~A.}\ \bibnamefont {Malomed}},\
  and\ \bibinfo {author} {\bibfnamefont {Z.}~\bibnamefont {Yan}},\ }\href
  {https://doi.org/10.1103/PhysRevE.110.014215} {\bibfield  {journal} {\bibinfo
   {journal} {Phys. Rev. E}\ }\textbf {\bibinfo {volume} {110}},\ \bibinfo
  {pages} {014215} (\bibinfo {year} {2024})}\BibitemShut {NoStop}%
\bibitem [{\citenamefont {Zu}\ \emph {et~al.}(2021)\citenamefont {Zu},
  \citenamefont {Gao},\ and\ \citenamefont {Yu}}]{zu2021}%
  \BibitemOpen
  \bibfield  {author} {\bibinfo {author} {\bibfnamefont {C.}~\bibnamefont
  {Zu}}, \bibinfo {author} {\bibfnamefont {Y.}~\bibnamefont {Gao}},\ and\
  \bibinfo {author} {\bibfnamefont {X.}~\bibnamefont {Yu}},\ }\href
  {https://doi.org/https://doi.org/10.1016/j.chaos.2021.110930} {\bibfield
  {journal} {\bibinfo  {journal} {Chaos Solit. Fractals}\ }\textbf {\bibinfo
  {volume} {147}},\ \bibinfo {pages} {110930} (\bibinfo {year}
  {2021})}\BibitemShut {NoStop}%
\bibitem [{\citenamefont {{El Allati}}\ \emph {et~al.}(2024)\citenamefont {{El
  Allati}}, \citenamefont {Bukbech}, \citenamefont {{El Anouz}},\ and\
  \citenamefont {{El Allali}}}]{elallati2024}%
  \BibitemOpen
  \bibfield  {author} {\bibinfo {author} {\bibfnamefont {A.}~\bibnamefont {{El
  Allati}}}, \bibinfo {author} {\bibfnamefont {S.}~\bibnamefont {Bukbech}},
  \bibinfo {author} {\bibfnamefont {K.}~\bibnamefont {{El Anouz}}},\ and\
  \bibinfo {author} {\bibfnamefont {Z.}~\bibnamefont {{El Allali}}},\ }\href
  {https://doi.org/https://doi.org/10.1016/j.chaos.2023.114446} {\bibfield
  {journal} {\bibinfo  {journal} {Chaos Solit. Fractals}\ }\textbf {\bibinfo
  {volume} {179}},\ \bibinfo {pages} {114446} (\bibinfo {year}
  {2024})}\BibitemShut {NoStop}%
\bibitem [{\citenamefont {Abdessamie}\ \emph {et~al.}(2025)\citenamefont
  {Abdessamie}, \citenamefont {Mansoura}, \citenamefont {Mansour},
  \citenamefont {Khadija}, \citenamefont {Ouchrif},\ and\ \citenamefont
  {El~Allati}}]{abdessamie2025}%
  \BibitemOpen
  \bibfield  {author} {\bibinfo {author} {\bibfnamefont {C.}~\bibnamefont
  {Abdessamie}}, \bibinfo {author} {\bibfnamefont {O.}~\bibnamefont
  {Mansoura}}, \bibinfo {author} {\bibfnamefont {M.}~\bibnamefont {Mansour}},
  \bibinfo {author} {\bibfnamefont {E.~A.}\ \bibnamefont {Khadija}}, \bibinfo
  {author} {\bibfnamefont {M.}~\bibnamefont {Ouchrif}},\ and\ \bibinfo {author}
  {\bibfnamefont {A.}~\bibnamefont {El~Allati}},\ }\href
  {https://doi.org/10.1088/1402-4896/addc50} {\bibfield  {journal} {\bibinfo
  {journal} {Phys. Scr.}\ }\textbf {\bibinfo {volume} {100}},\ \bibinfo {pages}
  {075105} (\bibinfo {year} {2025})}\BibitemShut {NoStop}%
\bibitem [{\citenamefont {Gabrick}\ \emph {et~al.}(2026)\citenamefont
  {Gabrick}, \citenamefont {Tsutsui}, \citenamefont {Cius}, \citenamefont
  {Castro}, \citenamefont {Andrade},\ and\ \citenamefont
  {Lenzi}}]{Gabrick2026}%
  \BibitemOpen
  \bibfield  {author} {\bibinfo {author} {\bibfnamefont {E.~C.}\ \bibnamefont
  {Gabrick}}, \bibinfo {author} {\bibfnamefont {T.~T.}\ \bibnamefont
  {Tsutsui}}, \bibinfo {author} {\bibfnamefont {D.}~\bibnamefont {Cius}},
  \bibinfo {author} {\bibfnamefont {A.~S. M.~D.}\ \bibnamefont {Castro}},
  \bibinfo {author} {\bibfnamefont {F.~M.}\ \bibnamefont {Andrade}},\ and\
  \bibinfo {author} {\bibfnamefont {E.~K.}\ \bibnamefont {Lenzi}},\ }\href
  {https://doi.org/10.1016/j.cnsns.2026.109912} {\bibfield  {journal} {\bibinfo
   {journal} {Commun. Nonlinear Sci. Numer. Simul.}\ }\textbf {\bibinfo
  {volume} {160}},\ \bibinfo {pages} {109912} (\bibinfo {year}
  {2026})}\BibitemShut {NoStop}%
\bibitem [{\citenamefont {Laskin}(2000{\natexlab{a}})}]{laskin:00a}%
  \BibitemOpen
  \bibfield  {author} {\bibinfo {author} {\bibfnamefont {N.}~\bibnamefont
  {Laskin}},\ }\href
  {https://doi.org/https://doi.org/10.1016/S0375-9601(00)00201-2} {\bibfield
  {journal} {\bibinfo  {journal} {Phys. Lett. A}\ }\textbf {\bibinfo {volume}
  {268}},\ \bibinfo {pages} {298} (\bibinfo {year}
  {2000}{\natexlab{a}})}\BibitemShut {NoStop}%
\bibitem [{\citenamefont {Laskin}(2000{\natexlab{b}})}]{laskin:00b}%
  \BibitemOpen
  \bibfield  {author} {\bibinfo {author} {\bibfnamefont {N.}~\bibnamefont
  {Laskin}},\ }\href {https://doi.org/https://doi.org/10.1103/PhysRevE.62.3135}
  {\bibfield  {journal} {\bibinfo  {journal} {Phys. Rev. E}\ }\textbf {\bibinfo
  {volume} {62}},\ \bibinfo {pages} {3135} (\bibinfo {year}
  {2000}{\natexlab{b}})}\BibitemShut {NoStop}%
\bibitem [{\citenamefont {Laskin}(2002)}]{laskin2002}%
  \BibitemOpen
  \bibfield  {author} {\bibinfo {author} {\bibfnamefont {N.}~\bibnamefont
  {Laskin}},\ }\href {https://doi.org/10.1103/PhysRevE.66.056108} {\bibfield
  {journal} {\bibinfo  {journal} {Phys. Rev. E}\ }\textbf {\bibinfo {volume}
  {66}},\ \bibinfo {pages} {056108} (\bibinfo {year} {2002})}\BibitemShut
  {NoStop}%
\bibitem [{\citenamefont {West}(2000)}]{west00}%
  \BibitemOpen
  \bibfield  {author} {\bibinfo {author} {\bibfnamefont {B.~J.}\ \bibnamefont
  {West}},\ }\href {https://doi.org/10.1021/jp993323u} {\bibfield  {journal}
  {\bibinfo  {journal} {J. Phys. Chem. B}\ }\textbf {\bibinfo {volume} {104}},\
  \bibinfo {pages} {3830} (\bibinfo {year} {2000})}\BibitemShut {NoStop}%
\bibitem [{\citenamefont {Feynman}\ and\ \citenamefont
  {Hibbs}(1965)}]{feynman1965}%
  \BibitemOpen
  \bibfield  {author} {\bibinfo {author} {\bibfnamefont {R.}~\bibnamefont
  {Feynman}}\ and\ \bibinfo {author} {\bibfnamefont {A.}~\bibnamefont
  {Hibbs}},\ }\href@noop {} {\emph {\bibinfo {title} {Quantum Mechanics and
  Path Integrals}}}\ (\bibinfo  {publisher} {McGraw-Hill},\ \bibinfo {address}
  {New York},\ \bibinfo {year} {1965})\BibitemShut {NoStop}%
\bibitem [{\citenamefont {Riascos}\ and\ \citenamefont
  {Mateos}(2015)}]{riascos2015}%
  \BibitemOpen
  \bibfield  {author} {\bibinfo {author} {\bibfnamefont {A.~P.}\ \bibnamefont
  {Riascos}}\ and\ \bibinfo {author} {\bibfnamefont {J.~L.}\ \bibnamefont
  {Mateos}},\ }\href {https://doi.org/10.1103/PhysRevE.92.052814} {\bibfield
  {journal} {\bibinfo  {journal} {Phys. Rev. E}\ }\textbf {\bibinfo {volume}
  {92}},\ \bibinfo {pages} {052814} (\bibinfo {year} {2015})}\BibitemShut
  {NoStop}%
\bibitem [{\citenamefont {Zhang}\ \emph {et~al.}(2017)\citenamefont {Zhang},
  \citenamefont {Zhang}, \citenamefont {Zhang}, \citenamefont {Ahmed},
  \citenamefont {Zhang}, \citenamefont {Li}, \citenamefont {Belić},\ and\
  \citenamefont {Xiao}}]{zhang2017}%
  \BibitemOpen
  \bibfield  {author} {\bibinfo {author} {\bibfnamefont {D.}~\bibnamefont
  {Zhang}}, \bibinfo {author} {\bibfnamefont {Y.}~\bibnamefont {Zhang}},
  \bibinfo {author} {\bibfnamefont {Z.}~\bibnamefont {Zhang}}, \bibinfo
  {author} {\bibfnamefont {N.}~\bibnamefont {Ahmed}}, \bibinfo {author}
  {\bibfnamefont {Y.}~\bibnamefont {Zhang}}, \bibinfo {author} {\bibfnamefont
  {F.}~\bibnamefont {Li}}, \bibinfo {author} {\bibfnamefont {M.~R.}\
  \bibnamefont {Belić}},\ and\ \bibinfo {author} {\bibfnamefont
  {M.}~\bibnamefont {Xiao}},\ }\href
  {https://doi.org/https://doi.org/10.1002/andp.201700149} {\bibfield
  {journal} {\bibinfo  {journal} {Ann. Phys. (Berl.)}\ }\textbf {\bibinfo
  {volume} {529}},\ \bibinfo {pages} {1700149} (\bibinfo {year}
  {2017})}\BibitemShut {NoStop}%
\bibitem [{\citenamefont {Naber}(2004)}]{naber:04}%
  \BibitemOpen
  \bibfield  {author} {\bibinfo {author} {\bibfnamefont {M.}~\bibnamefont
  {Naber}},\ }\href {https://doi.org/https://doi.org/10.1063/1.1769611}
  {\bibfield  {journal} {\bibinfo  {journal} {J. Math. Phys.}\ }\textbf
  {\bibinfo {volume} {45}},\ \bibinfo {pages} {3339} (\bibinfo {year}
  {2004})}\BibitemShut {NoStop}%
\bibitem [{\citenamefont {Lenzi}\ \emph {et~al.}(2013)\citenamefont {Lenzi},
  \citenamefont {Ribeiro}, \citenamefont {dos Santos}, \citenamefont
  {Rossato},\ and\ \citenamefont {Mendes}}]{lenzi2013time}%
  \BibitemOpen
  \bibfield  {author} {\bibinfo {author} {\bibfnamefont {E.~K.}\ \bibnamefont
  {Lenzi}}, \bibinfo {author} {\bibfnamefont {H.~V.}\ \bibnamefont {Ribeiro}},
  \bibinfo {author} {\bibfnamefont {M.~A.~F.}\ \bibnamefont {dos Santos}},
  \bibinfo {author} {\bibfnamefont {R.}~\bibnamefont {Rossato}},\ and\ \bibinfo
  {author} {\bibfnamefont {R.~S.}\ \bibnamefont {Mendes}},\ }\href
  {https://doi.org/https://doi.org/10.1063/1.4819253} {\bibfield  {journal}
  {\bibinfo  {journal} {J. Math. Phys.}\ }\textbf {\bibinfo {volume} {54}},\
  \bibinfo {pages} {082107} (\bibinfo {year} {2013})}\BibitemShut {NoStop}%
\bibitem [{\citenamefont {Gabrick}\ \emph {et~al.}(2023)\citenamefont
  {Gabrick}, \citenamefont {Sayari}, \citenamefont {{de Castro}}, \citenamefont
  {Trobia}, \citenamefont {Batista},\ and\ \citenamefont
  {Lenzi}}]{gabrick2023}%
  \BibitemOpen
  \bibfield  {author} {\bibinfo {author} {\bibfnamefont {E.}~\bibnamefont
  {Gabrick}}, \bibinfo {author} {\bibfnamefont {E.}~\bibnamefont {Sayari}},
  \bibinfo {author} {\bibfnamefont {A.}~\bibnamefont {{de Castro}}}, \bibinfo
  {author} {\bibfnamefont {J.}~\bibnamefont {Trobia}}, \bibinfo {author}
  {\bibfnamefont {A.}~\bibnamefont {Batista}},\ and\ \bibinfo {author}
  {\bibfnamefont {E.}~\bibnamefont {Lenzi}},\ }\href
  {https://doi.org/https://doi.org/10.1016/j.cnsns.2023.107275} {\bibfield
  {journal} {\bibinfo  {journal} {Commun. Nonlinear Sci. Numer. Simul.}\
  }\textbf {\bibinfo {volume} {123}},\ \bibinfo {pages} {107275} (\bibinfo
  {year} {2023})}\BibitemShut {NoStop}%
\bibitem [{\citenamefont {Iomin}(2009)}]{iomin2009fractional}%
  \BibitemOpen
  \bibfield  {author} {\bibinfo {author} {\bibfnamefont {A.}~\bibnamefont
  {Iomin}},\ }\href
  {https://doi.org/https://doi.org/10.1103/PhysRevE.80.022103} {\bibfield
  {journal} {\bibinfo  {journal} {Phys. Rev. E}\ }\textbf {\bibinfo {volume}
  {80}},\ \bibinfo {pages} {022103} (\bibinfo {year} {2009})}\BibitemShut
  {NoStop}%
\bibitem [{\citenamefont {Iomin}(2024)}]{iomin2024}%
  \BibitemOpen
  \bibfield  {author} {\bibinfo {author} {\bibfnamefont {A.}~\bibnamefont
  {Iomin}},\ }\href {https://doi.org/10.1063/5.0226335} {\bibfield  {journal}
  {\bibinfo  {journal} {Chaos}\ }\textbf {\bibinfo {volume} {34}},\ \bibinfo
  {pages} {093135} (\bibinfo {year} {2024})}\BibitemShut {NoStop}%
\bibitem [{\citenamefont {Zu}\ and\ \citenamefont {Yu}(2025)}]{Zu2025}%
  \BibitemOpen
  \bibfield  {author} {\bibinfo {author} {\bibfnamefont {C.}~\bibnamefont
  {Zu}}\ and\ \bibinfo {author} {\bibfnamefont {X.}~\bibnamefont {Yu}},\ }\href
  {https://doi.org/10.1063/5.0253816} {\bibfield  {journal} {\bibinfo
  {journal} {J. Chem. Phys.}\ }\textbf {\bibinfo {volume} {162}},\ \bibinfo
  {pages} {3830} (\bibinfo {year} {2025})}\BibitemShut {NoStop}%
\bibitem [{\citenamefont {Breuer}\ and\ \citenamefont
  {Petruccione}(2007)}]{breuer2007}%
  \BibitemOpen
  \bibfield  {author} {\bibinfo {author} {\bibfnamefont {H.}~\bibnamefont
  {Breuer}}\ and\ \bibinfo {author} {\bibfnamefont {F.}~\bibnamefont
  {Petruccione}},\ }\href@noop {} {\emph {\bibinfo {title} {The Theory of Open
  Quantum Systems}}}\ (\bibinfo  {publisher} {OUP Oxford},\ \bibinfo {year}
  {2007})\BibitemShut {NoStop}%
\bibitem [{\citenamefont {Tarasov}(2010)}]{tarasov2010}%
  \BibitemOpen
  \bibfield  {author} {\bibinfo {author} {\bibfnamefont {V.~E.}\ \bibnamefont
  {Tarasov}},\ }\bibinfo {title} {Fractional {D}ynamics of {O}pen {Q}uantum
  {S}ystems},\ in\ \href {https://doi.org/10.1007/978-3-642-14003-7_20} {\emph
  {\bibinfo {booktitle} {Fractional Dynamics: Applications of Fractional
  Calculus to Dynamics of Particles, Fields and Media}}}\ (\bibinfo
  {publisher} {Springer Berlin Heidelberg},\ \bibinfo {address} {Berlin,
  Heidelberg},\ \bibinfo {year} {2010})\ pp.\ \bibinfo {pages}
  {467--490}\BibitemShut {NoStop}%
\bibitem [{\citenamefont {Tarasov}(2012)}]{tarasov2012}%
  \BibitemOpen
  \bibfield  {author} {\bibinfo {author} {\bibfnamefont {V.~E.}\ \bibnamefont
  {Tarasov}},\ }\href
  {https://doi.org/https://doi.org/10.1016/j.aop.2012.02.011} {\bibfield
  {journal} {\bibinfo  {journal} {Ann. Phys.}\ }\textbf {\bibinfo {volume}
  {327}},\ \bibinfo {pages} {1719} (\bibinfo {year} {2012})}\BibitemShut
  {NoStop}%
\bibitem [{\citenamefont {Wang}\ and\ \citenamefont {Xu}(2007)}]{wang2007}%
  \BibitemOpen
  \bibfield  {author} {\bibinfo {author} {\bibfnamefont {S.}~\bibnamefont
  {Wang}}\ and\ \bibinfo {author} {\bibfnamefont {M.}~\bibnamefont {Xu}},\
  }\href {https://doi.org/https://doi.org/10.1063/1.2716203} {\bibfield
  {journal} {\bibinfo  {journal} {J. Math. Phys.}\ }\textbf {\bibinfo {volume}
  {48}},\ \bibinfo {pages} {043502} (\bibinfo {year} {2007})}\BibitemShut
  {NoStop}%
\bibitem [{\citenamefont {Dong}\ and\ \citenamefont {Xu}(2008)}]{dong2008}%
  \BibitemOpen
  \bibfield  {author} {\bibinfo {author} {\bibfnamefont {J.}~\bibnamefont
  {Dong}}\ and\ \bibinfo {author} {\bibfnamefont {M.}~\bibnamefont {Xu}},\
  }\href {https://doi.org/https://doi.org/10.1016/j.jmaa.2008.03.061}
  {\bibfield  {journal} {\bibinfo  {journal} {J. Math. Anal. Appl.}\ }\textbf
  {\bibinfo {volume} {344}},\ \bibinfo {pages} {1005} (\bibinfo {year}
  {2008})}\BibitemShut {NoStop}%
\bibitem [{\citenamefont {Lenzi}\ \emph {et~al.}(2023)\citenamefont {Lenzi},
  \citenamefont {Gabrick}, \citenamefont {Sayari}, \citenamefont {de~Castro},
  \citenamefont {Trobia},\ and\ \citenamefont {Batista}}]{lenzi2023}%
  \BibitemOpen
  \bibfield  {author} {\bibinfo {author} {\bibfnamefont {E.~K.}\ \bibnamefont
  {Lenzi}}, \bibinfo {author} {\bibfnamefont {E.~C.}\ \bibnamefont {Gabrick}},
  \bibinfo {author} {\bibfnamefont {E.}~\bibnamefont {Sayari}}, \bibinfo
  {author} {\bibfnamefont {A.~S.~M.}\ \bibnamefont {de~Castro}}, \bibinfo
  {author} {\bibfnamefont {J.}~\bibnamefont {Trobia}},\ and\ \bibinfo {author}
  {\bibfnamefont {A.~M.}\ \bibnamefont {Batista}},\ }\href
  {https://doi.org/10.3390/quantum5020029} {\bibfield  {journal} {\bibinfo
  {journal} {Quantum Rep.}\ }\textbf {\bibinfo {volume} {5}},\ \bibinfo {pages}
  {442} (\bibinfo {year} {2023})}\BibitemShut {NoStop}%
\bibitem [{\citenamefont {Laskin}(2017)}]{laskin2017time}%
  \BibitemOpen
  \bibfield  {author} {\bibinfo {author} {\bibfnamefont {N.}~\bibnamefont
  {Laskin}},\ }\href
  {https://doi.org/https://doi.org/10.1016/j.chaos.2017.04.010} {\bibfield
  {journal} {\bibinfo  {journal} {Chaos Solit. Fractals}\ }\textbf {\bibinfo
  {volume} {102}},\ \bibinfo {pages} {16} (\bibinfo {year} {2017})}\BibitemShut
  {NoStop}%
\bibitem [{\citenamefont {Ertik}\ \emph {et~al.}(2010)\citenamefont {Ertik},
  \citenamefont {Demirhan}, \citenamefont {Şirin},\ and\ \citenamefont
  {Büyükkılıç}}]{ertik2010}%
  \BibitemOpen
  \bibfield  {author} {\bibinfo {author} {\bibfnamefont {H.}~\bibnamefont
  {Ertik}}, \bibinfo {author} {\bibfnamefont {D.}~\bibnamefont {Demirhan}},
  \bibinfo {author} {\bibfnamefont {H.}~\bibnamefont {Şirin}},\ and\ \bibinfo
  {author} {\bibfnamefont {F.}~\bibnamefont {Büyükkılıç}},\ }\href
  {https://doi.org/10.1063/1.3464492} {\bibfield  {journal} {\bibinfo
  {journal} {J. Math. Phys.}\ }\textbf {\bibinfo {volume} {51}},\ \bibinfo
  {pages} {082102} (\bibinfo {year} {2010})}\BibitemShut {NoStop}%
\bibitem [{\citenamefont {Iomin}(2019{\natexlab{a}})}]{iomin2019fractional}%
  \BibitemOpen
  \bibfield  {author} {\bibinfo {author} {\bibfnamefont {A.}~\bibnamefont
  {Iomin}},\ }\href
  {https://doi.org/https://doi.org/10.1016/j.csfx.2018.100001} {\bibfield
  {journal} {\bibinfo  {journal} {Chaos Solit. Fractals: X}\ }\textbf {\bibinfo
  {volume} {1}},\ \bibinfo {pages} {100001} (\bibinfo {year}
  {2019}{\natexlab{a}})}\BibitemShut {NoStop}%
\bibitem [{\citenamefont {Stone}(1932)}]{stone1932}%
  \BibitemOpen
  \bibfield  {author} {\bibinfo {author} {\bibfnamefont {M.~H.}\ \bibnamefont
  {Stone}},\ }\href {https://doi.org/10.2307/1968538} {\bibfield  {journal}
  {\bibinfo  {journal} {Ann. Math.}\ }\textbf {\bibinfo {volume} {33}},\
  \bibinfo {pages} {643} (\bibinfo {year} {1932})}\BibitemShut {NoStop}%
\bibitem [{\citenamefont {Cius}\ \emph {et~al.}(2022)\citenamefont {Cius},
  \citenamefont {Menon}, \citenamefont {dos Santos}, \citenamefont
  {de~Castro},\ and\ \citenamefont {Andrade}}]{cius22frac}%
  \BibitemOpen
  \bibfield  {author} {\bibinfo {author} {\bibfnamefont {D.}~\bibnamefont
  {Cius}}, \bibinfo {author} {\bibfnamefont {L.}~\bibnamefont {Menon}},
  \bibinfo {author} {\bibfnamefont {M.~A.~F.}\ \bibnamefont {dos Santos}},
  \bibinfo {author} {\bibfnamefont {A.~S.~M.}\ \bibnamefont {de~Castro}},\ and\
  \bibinfo {author} {\bibfnamefont {F.~M.}\ \bibnamefont {Andrade}},\ }\href
  {https://doi.org/10.1103/PhysRevE.106.054126} {\bibfield  {journal} {\bibinfo
   {journal} {Phys. Rev. E}\ }\textbf {\bibinfo {volume} {106}},\ \bibinfo
  {pages} {054126} (\bibinfo {year} {2022})}\BibitemShut {NoStop}%
\bibitem [{\citenamefont {Jaynes}\ and\ \citenamefont
  {Cummings}(1963)}]{jaynes1963}%
  \BibitemOpen
  \bibfield  {author} {\bibinfo {author} {\bibfnamefont {E.}~\bibnamefont
  {Jaynes}}\ and\ \bibinfo {author} {\bibfnamefont {F.}~\bibnamefont
  {Cummings}},\ }\href {https://doi.org/10.1109/PROC.1963.1664} {\bibfield
  {journal} {\bibinfo  {journal} {Proc. IEEE}\ }\textbf {\bibinfo {volume}
  {51}},\ \bibinfo {pages} {89} (\bibinfo {year} {1963})}\BibitemShut {NoStop}%
\bibitem [{\citenamefont {Uhdre}\ \emph {et~al.}(2022)\citenamefont {Uhdre},
  \citenamefont {Cius},\ and\ \citenamefont {Andrade}}]{UHDRE2022}%
  \BibitemOpen
  \bibfield  {author} {\bibinfo {author} {\bibfnamefont {G.~M.}\ \bibnamefont
  {Uhdre}}, \bibinfo {author} {\bibfnamefont {D.}~\bibnamefont {Cius}},\ and\
  \bibinfo {author} {\bibfnamefont {F.~M.}\ \bibnamefont {Andrade}},\ }\href
  {https://doi.org/10.1103/PhysRevA.105.013703} {\bibfield  {journal} {\bibinfo
   {journal} {Phys. Rev. A}\ }\textbf {\bibinfo {volume} {105}},\ \bibinfo
  {pages} {013703} (\bibinfo {year} {2022})}\BibitemShut {NoStop}%
\bibitem [{\citenamefont {Vidiella‐Barranco}\ \emph
  {et~al.}(2025)\citenamefont {Vidiella‐Barranco}, \citenamefont
  {{Magalh{\~{a}}es de Castro}}, \citenamefont {Sergi}, \citenamefont
  {Roversi}, \citenamefont {Messina},\ and\ \citenamefont
  {Migliore}}]{VidiellaBarranco2025}%
  \BibitemOpen
  \bibfield  {author} {\bibinfo {author} {\bibfnamefont {A.}~\bibnamefont
  {Vidiella‐Barranco}}, \bibinfo {author} {\bibfnamefont {A.~S.}\
  \bibnamefont {{Magalh{\~{a}}es de Castro}}}, \bibinfo {author} {\bibfnamefont
  {A.}~\bibnamefont {Sergi}}, \bibinfo {author} {\bibfnamefont {J.~A.}\
  \bibnamefont {Roversi}}, \bibinfo {author} {\bibfnamefont {A.}~\bibnamefont
  {Messina}},\ and\ \bibinfo {author} {\bibfnamefont {A.}~\bibnamefont
  {Migliore}},\ }\href {https://doi.org/10.1002/andp.202500148} {\bibfield
  {journal} {\bibinfo  {journal} {Ann. Phys.}\ }\textbf {\bibinfo {volume}
  {537}},\ \bibinfo {pages} {e00148} (\bibinfo {year} {2025})}\BibitemShut
  {NoStop}%
\bibitem [{\citenamefont {Arrazola}\ \emph {et~al.}(2025)\citenamefont
  {Arrazola}, \citenamefont {Bertet}, \citenamefont {Chu},\ and\ \citenamefont
  {Rabl}}]{Arrazola2025}%
  \BibitemOpen
  \bibfield  {author} {\bibinfo {author} {\bibfnamefont {I.}~\bibnamefont
  {Arrazola}}, \bibinfo {author} {\bibfnamefont {P.}~\bibnamefont {Bertet}},
  \bibinfo {author} {\bibfnamefont {Y.}~\bibnamefont {Chu}},\ and\ \bibinfo
  {author} {\bibfnamefont {P.}~\bibnamefont {Rabl}},\ }\href
  {https://doi.org/10.1038/s41534-025-01143-5} {\bibfield  {journal} {\bibinfo
  {journal} {npj Quantum Inf.}\ }\textbf {\bibinfo {volume} {11}},\ \bibinfo
  {pages} {1} (\bibinfo {year} {2025})},\ \Eprint
  {https://arxiv.org/abs/2505.02929} {2505.02929} \BibitemShut {NoStop}%
\bibitem [{\citenamefont {Zhou}\ \emph {et~al.}(2025)\citenamefont {Zhou},
  \citenamefont {Liu}, \citenamefont {Su}, \citenamefont {Zhang}, \citenamefont
  {Wu}, \citenamefont {Chen}, \citenamefont {Shi}, \citenamefont {Shen},\ and\
  \citenamefont {Yang}}]{Zhou2025a}%
  \BibitemOpen
  \bibfield  {author} {\bibinfo {author} {\bibfnamefont {Y.~H.}\ \bibnamefont
  {Zhou}}, \bibinfo {author} {\bibfnamefont {T.}~\bibnamefont {Liu}}, \bibinfo
  {author} {\bibfnamefont {Q.~P.}\ \bibnamefont {Su}}, \bibinfo {author}
  {\bibfnamefont {X.~Y.}\ \bibnamefont {Zhang}}, \bibinfo {author}
  {\bibfnamefont {Q.~C.}\ \bibnamefont {Wu}}, \bibinfo {author} {\bibfnamefont
  {D.~X.}\ \bibnamefont {Chen}}, \bibinfo {author} {\bibfnamefont {Z.~C.}\
  \bibnamefont {Shi}}, \bibinfo {author} {\bibfnamefont {H.~Z.}\ \bibnamefont
  {Shen}},\ and\ \bibinfo {author} {\bibfnamefont {C.~P.}\ \bibnamefont
  {Yang}},\ }\href {https://doi.org/10.1103/PhysRevLett.134.183601} {\bibfield
  {journal} {\bibinfo  {journal} {Phys. Rev. Lett.}\ }\textbf {\bibinfo
  {volume} {134}},\ \bibinfo {pages} {183601} (\bibinfo {year}
  {2025})}\BibitemShut {NoStop}%
\bibitem [{\citenamefont {Tsutsui}\ \emph {et~al.}(2026)\citenamefont
  {Tsutsui}, \citenamefont {Cius}, \citenamefont {Vidiella-Barranco},
  \citenamefont {de~Castro},\ and\ \citenamefont {Andrade}}]{Tsutsui2026}%
  \BibitemOpen
  \bibfield  {author} {\bibinfo {author} {\bibfnamefont {T.~T.}\ \bibnamefont
  {Tsutsui}}, \bibinfo {author} {\bibfnamefont {D.}~\bibnamefont {Cius}},
  \bibinfo {author} {\bibfnamefont {A.}~\bibnamefont {Vidiella-Barranco}},
  \bibinfo {author} {\bibfnamefont {A.~S.~M.}\ \bibnamefont {de~Castro}},\ and\
  \bibinfo {author} {\bibfnamefont {F.~M.}\ \bibnamefont {Andrade}},\ }\href
  {https://doi.org/10.1007/s13538-025-01949-w} {\bibfield  {journal} {\bibinfo
  {journal} {Brazilian J. Phys.}\ }\textbf {\bibinfo {volume} {56}},\ \bibinfo
  {pages} {21} (\bibinfo {year} {2026})}\BibitemShut {NoStop}%
\bibitem [{\citenamefont {Cius}(2025)}]{Cius2024}%
  \BibitemOpen
  \bibfield  {author} {\bibinfo {author} {\bibfnamefont {D.}~\bibnamefont
  {Cius}},\ }\href {https://doi.org/10.1103/PhysRevE.111.024110} {\bibfield
  {journal} {\bibinfo  {journal} {Phys. Rev. E}\ }\textbf {\bibinfo {volume}
  {111}},\ \bibinfo {pages} {024110} (\bibinfo {year} {2025})}\BibitemShut
  {NoStop}%
\bibitem [{\citenamefont {Puri}\ and\ \citenamefont
  {Bullough}(1988)}]{Puri1988}%
  \BibitemOpen
  \bibfield  {author} {\bibinfo {author} {\bibfnamefont {R.~R.}\ \bibnamefont
  {Puri}}\ and\ \bibinfo {author} {\bibfnamefont {R.~K.}\ \bibnamefont
  {Bullough}},\ }\href {https://doi.org/10.1364/JOSAB.5.002021} {\bibfield
  {journal} {\bibinfo  {journal} {J. Opt. Soc. Am. B}\ }\textbf {\bibinfo
  {volume} {5}},\ \bibinfo {pages} {2021} (\bibinfo {year} {1988})}\BibitemShut
  {NoStop}%
\bibitem [{\citenamefont {Rabi}(1936)}]{Rabi1936}%
  \BibitemOpen
  \bibfield  {author} {\bibinfo {author} {\bibfnamefont {I.~I.}\ \bibnamefont
  {Rabi}},\ }\href {https://doi.org/10.1103/PhysRev.49.324} {\bibfield
  {journal} {\bibinfo  {journal} {Phys. Rev.}\ }\textbf {\bibinfo {volume}
  {49}},\ \bibinfo {pages} {324} (\bibinfo {year} {1936})}\BibitemShut
  {NoStop}%
\bibitem [{\citenamefont {Rabi}(1937)}]{Rabi1937}%
  \BibitemOpen
  \bibfield  {author} {\bibinfo {author} {\bibfnamefont {I.~I.}\ \bibnamefont
  {Rabi}},\ }\href {https://doi.org/10.1103/PhysRev.51.652} {\bibfield
  {journal} {\bibinfo  {journal} {Phys. Rev.}\ }\textbf {\bibinfo {volume}
  {51}},\ \bibinfo {pages} {652} (\bibinfo {year} {1937})}\BibitemShut
  {NoStop}%
\bibitem [{\citenamefont {Klimov}\ and\ \citenamefont
  {Chumakov}(2009)}]{Klimov2009}%
  \BibitemOpen
  \bibfield  {author} {\bibinfo {author} {\bibfnamefont {A.~B.}\ \bibnamefont
  {Klimov}}\ and\ \bibinfo {author} {\bibfnamefont {S.~M.}\ \bibnamefont
  {Chumakov}},\ }\href {https://doi.org/10.1002/9783527624003} {\emph {\bibinfo
  {title} {{A Group Theoretical Approach to Quantum Optics: Models of Atom
  Field Interactions }}}}\ (\bibinfo  {publisher} {John Wiley \& Sons},\
  \bibinfo {address} {Weinheim},\ \bibinfo {year} {2009})\BibitemShut {NoStop}%
\bibitem [{\citenamefont {Cantuba}(2024)}]{Cantuba2024}%
  \BibitemOpen
  \bibfield  {author} {\bibinfo {author} {\bibfnamefont {R.~R.~S.}\
  \bibnamefont {Cantuba}},\ }\href {https://doi.org/10.24330/ieja.1326849}
  {\bibfield  {journal} {\bibinfo  {journal} {Int. Electron. J. Algebra}\
  }\textbf {\bibinfo {volume} {35}},\ \bibinfo {pages} {32–60} (\bibinfo
  {year} {2024})}\BibitemShut {NoStop}%
\bibitem [{\citenamefont {Larson}\ and\ \citenamefont
  {Mavrogordatos}(2021)}]{larson2021}%
  \BibitemOpen
  \bibfield  {author} {\bibinfo {author} {\bibfnamefont {J.}~\bibnamefont
  {Larson}}\ and\ \bibinfo {author} {\bibfnamefont {T.}~\bibnamefont
  {Mavrogordatos}},\ }\href {https://doi.org/10.1088/978-0-7503-3447-1} {\emph
  {\bibinfo {title} {The Jaynes–Cummings Model and Its Descendants}}},\
  2053-2563\ (\bibinfo  {publisher} {IOP Publishing},\ \bibinfo {year}
  {2021})\BibitemShut {NoStop}%
\bibitem [{\citenamefont {Iomin}(2019{\natexlab{b}})}]{iomin2019app}%
  \BibitemOpen
  \bibfield  {author} {\bibinfo {author} {\bibfnamefont {A.}~\bibnamefont
  {Iomin}},\ }\bibinfo {title} {Fractional time quantum mechanics},\ in\ \href
  {https://doi.org/doi:10.1515/9783110571721-013} {\emph {\bibinfo {booktitle}
  {Volume 5 Applications in Physics, Part B}}},\ \bibinfo {editor} {edited by\
  \bibinfo {editor} {\bibfnamefont {V.~E.}\ \bibnamefont {Tarasov}}}\ (\bibinfo
   {publisher} {De Gruyter},\ \bibinfo {address} {Berlin, Boston},\ \bibinfo
  {year} {2019})\ pp.\ \bibinfo {pages} {299--316}\BibitemShut {NoStop}%
\bibitem [{\citenamefont {Fring}\ and\ \citenamefont
  {Moussa}(2016)}]{fring:16a}%
  \BibitemOpen
  \bibfield  {author} {\bibinfo {author} {\bibfnamefont {A.}~\bibnamefont
  {Fring}}\ and\ \bibinfo {author} {\bibfnamefont {M.~H.~Y.}\ \bibnamefont
  {Moussa}},\ }\href
  {https://doi.org/https://doi.org/10.1103/PhysRevA.93.042114} {\bibfield
  {journal} {\bibinfo  {journal} {Phys. Rev. A}\ }\textbf {\bibinfo {volume}
  {93}},\ \bibinfo {pages} {042114} (\bibinfo {year} {2016})}\BibitemShut
  {NoStop}%
\bibitem [{\citenamefont {Fring}\ and\ \citenamefont {Frith}(2017)}]{fring:17}%
  \BibitemOpen
  \bibfield  {author} {\bibinfo {author} {\bibfnamefont {A.}~\bibnamefont
  {Fring}}\ and\ \bibinfo {author} {\bibfnamefont {T.}~\bibnamefont {Frith}},\
  }\href {https://doi.org/https://doi.org/10.1016/j.physleta.2017.05.041}
  {\bibfield  {journal} {\bibinfo  {journal} {Phys. Lett. A}\ }\textbf
  {\bibinfo {volume} {381}},\ \bibinfo {pages} {2318} (\bibinfo {year}
  {2017})}\BibitemShut {NoStop}%
\bibitem [{\citenamefont {Gerry}\ and\ \citenamefont
  {Knight}(2004)}]{GERRY2005}%
  \BibitemOpen
  \bibfield  {author} {\bibinfo {author} {\bibfnamefont {C.}~\bibnamefont
  {Gerry}}\ and\ \bibinfo {author} {\bibfnamefont {P.}~\bibnamefont {Knight}},\
  }\href {https://doi.org/10.1017/CBO9780511791239} {\emph {\bibinfo {title}
  {{Introductory Quantum Optics}}}},\ \bibinfo {edition} {2nd}\ ed.\ (\bibinfo
  {publisher} {Cambridge University Press},\ \bibinfo {address} {Cambridge},\
  \bibinfo {year} {2004})\BibitemShut {NoStop}%
\bibitem [{\citenamefont {Nielsen}\ and\ \citenamefont
  {Chuang}(2010)}]{Nielsen2010}%
  \BibitemOpen
  \bibfield  {author} {\bibinfo {author} {\bibfnamefont {M.~A.}\ \bibnamefont
  {Nielsen}}\ and\ \bibinfo {author} {\bibfnamefont {I.~L.}\ \bibnamefont
  {Chuang}},\ }\href@noop {} {\emph {\bibinfo {title} {{Quantum Computation and
  Quantum Information: 10th Anniversary Edition}}}},\ \bibinfo {edition}
  {10th}\ ed.\ (\bibinfo  {publisher} {Cambridge University Press},\ \bibinfo
  {address} {Cambridge},\ \bibinfo {year} {2010})\BibitemShut {NoStop}%
\bibitem [{\citenamefont {{C. Zhang}}\ and\ \citenamefont {{J.
  Jing.}}(2024)}]{Zhang2024}%
  \BibitemOpen
  \bibfield  {author} {\bibinfo {author} {\bibnamefont {{C. Zhang}}}\ and\
  \bibinfo {author} {\bibnamefont {{J. Jing.}}},\ }\href
  {https://doi.org/10.1103/PhysRevA.110.042421} {\bibfield  {journal} {\bibinfo
   {journal} {Phys. Rev. A}\ }\textbf {\bibinfo {volume} {110}},\ \bibinfo
  {pages} {042421} (\bibinfo {year} {2024})}\BibitemShut {NoStop}%
\bibitem [{\citenamefont {Rempe}\ \emph {et~al.}(1987)\citenamefont {Rempe},
  \citenamefont {Walther},\ and\ \citenamefont {Klein}}]{REMPE1987}%
  \BibitemOpen
  \bibfield  {author} {\bibinfo {author} {\bibfnamefont {G.}~\bibnamefont
  {Rempe}}, \bibinfo {author} {\bibfnamefont {H.}~\bibnamefont {Walther}},\
  and\ \bibinfo {author} {\bibfnamefont {N.}~\bibnamefont {Klein}},\ }\href
  {https://doi.org/10.1103/PhysRevLett.58.353} {\bibfield  {journal} {\bibinfo
  {journal} {Phys. Rev. Lett.}\ }\textbf {\bibinfo {volume} {58}},\ \bibinfo
  {pages} {353} (\bibinfo {year} {1987})}\BibitemShut {NoStop}%
\bibitem [{\citenamefont {Brune}\ \emph {et~al.}(1996)\citenamefont {Brune},
  \citenamefont {Schmidt-Kaler}, \citenamefont {Maali}, \citenamefont {Dreyer},
  \citenamefont {Hagley}, \citenamefont {Raimond},\ and\ \citenamefont
  {Haroche}}]{Brune1996}%
  \BibitemOpen
  \bibfield  {author} {\bibinfo {author} {\bibfnamefont {M.}~\bibnamefont
  {Brune}}, \bibinfo {author} {\bibfnamefont {F.}~\bibnamefont
  {Schmidt-Kaler}}, \bibinfo {author} {\bibfnamefont {A.}~\bibnamefont
  {Maali}}, \bibinfo {author} {\bibfnamefont {J.}~\bibnamefont {Dreyer}},
  \bibinfo {author} {\bibfnamefont {E.}~\bibnamefont {Hagley}}, \bibinfo
  {author} {\bibfnamefont {J.~M.}\ \bibnamefont {Raimond}},\ and\ \bibinfo
  {author} {\bibfnamefont {S.}~\bibnamefont {Haroche}},\ }\href
  {https://doi.org/10.1103/PhysRevLett.76.1800} {\bibfield  {journal} {\bibinfo
   {journal} {Phys. Rev. Lett.}\ }\textbf {\bibinfo {volume} {76}},\ \bibinfo
  {pages} {1800} (\bibinfo {year} {1996})}\BibitemShut {NoStop}%
\bibitem [{\citenamefont {Arroyo-Correa}\ and\ \citenamefont
  {Sanchez-Mondragon}(1990)}]{Arroyo-Correa1990}%
  \BibitemOpen
  \bibfield  {author} {\bibinfo {author} {\bibfnamefont {G.}~\bibnamefont
  {Arroyo-Correa}}\ and\ \bibinfo {author} {\bibfnamefont {J.~J.}\ \bibnamefont
  {Sanchez-Mondragon}},\ }\href {https://doi.org/10.1088/0954-8998/2/6/001}
  {\bibfield  {journal} {\bibinfo  {journal} {Quant. Optics}\ }\textbf
  {\bibinfo {volume} {2}},\ \bibinfo {pages} {409} (\bibinfo {year}
  {1990})}\BibitemShut {NoStop}%
\bibitem [{\citenamefont {Wei}\ \emph {et~al.}(2024)\citenamefont {Wei},
  \citenamefont {Liu}, \citenamefont {Li}, \citenamefont {Wan}, \citenamefont
  {Qin}, \citenamefont {Wen},\ and\ \citenamefont {Gao}}]{Wei2024}%
  \BibitemOpen
  \bibfield  {author} {\bibinfo {author} {\bibfnamefont {D.}~\bibnamefont
  {Wei}}, \bibinfo {author} {\bibfnamefont {H.}~\bibnamefont {Liu}}, \bibinfo
  {author} {\bibfnamefont {Y.}~\bibnamefont {Li}}, \bibinfo {author}
  {\bibfnamefont {L.}~\bibnamefont {Wan}}, \bibinfo {author} {\bibfnamefont
  {S.}~\bibnamefont {Qin}}, \bibinfo {author} {\bibfnamefont {Q.}~\bibnamefont
  {Wen}},\ and\ \bibinfo {author} {\bibfnamefont {F.}~\bibnamefont {Gao}},\
  }\href {https://doi.org/10.1016/j.chaos.2024.114816} {\bibfield  {journal}
  {\bibinfo  {journal} {Chaos Solit. Fractals}\ }\textbf {\bibinfo {volume}
  {182}},\ \bibinfo {pages} {114816} (\bibinfo {year} {2024})}\BibitemShut
  {NoStop}%
\bibitem [{\citenamefont {Wootters}(1998)}]{Wootters1998}%
  \BibitemOpen
  \bibfield  {author} {\bibinfo {author} {\bibfnamefont {W.~K.}\ \bibnamefont
  {Wootters}},\ }\href {https://doi.org/10.1103/PhysRevLett.80.2245} {\bibfield
   {journal} {\bibinfo  {journal} {Phys. Rev. Lett.}\ }\textbf {\bibinfo
  {volume} {80}},\ \bibinfo {pages} {2245} (\bibinfo {year}
  {1998})}\BibitemShut {NoStop}%
\bibitem [{\citenamefont {Gambetta}\ \emph {et~al.}(2011)\citenamefont
  {Gambetta}, \citenamefont {Houck},\ and\ \citenamefont
  {Blais}}]{Gambetta2011}%
  \BibitemOpen
  \bibfield  {author} {\bibinfo {author} {\bibfnamefont {J.~M.}\ \bibnamefont
  {Gambetta}}, \bibinfo {author} {\bibfnamefont {A.~A.}\ \bibnamefont
  {Houck}},\ and\ \bibinfo {author} {\bibfnamefont {A.}~\bibnamefont {Blais}},\
  }\href {https://doi.org/10.1103/PhysRevLett.106.030502} {\bibfield  {journal}
  {\bibinfo  {journal} {Phys. Rev. Lett.}\ }\textbf {\bibinfo {volume} {106}},\
  \bibinfo {pages} {030502} (\bibinfo {year} {2011})}\BibitemShut {NoStop}%
\bibitem [{\citenamefont {Srinivasan}\ \emph {et~al.}(2011)\citenamefont
  {Srinivasan}, \citenamefont {Hoffman}, \citenamefont {Gambetta},\ and\
  \citenamefont {Houck}}]{Srinivasan2011}%
  \BibitemOpen
  \bibfield  {author} {\bibinfo {author} {\bibfnamefont {S.~J.}\ \bibnamefont
  {Srinivasan}}, \bibinfo {author} {\bibfnamefont {A.~J.}\ \bibnamefont
  {Hoffman}}, \bibinfo {author} {\bibfnamefont {J.~M.}\ \bibnamefont
  {Gambetta}},\ and\ \bibinfo {author} {\bibfnamefont {A.~A.}\ \bibnamefont
  {Houck}},\ }\href {https://doi.org/10.1103/PhysRevLett.106.083601} {\bibfield
   {journal} {\bibinfo  {journal} {Phys. Rev. Lett.}\ }\textbf {\bibinfo
  {volume} {106}},\ \bibinfo {pages} {083601} (\bibinfo {year}
  {2011})}\BibitemShut {NoStop}%
\bibitem [{\citenamefont {Yin}\ \emph {et~al.}(2013)\citenamefont {Yin},
  \citenamefont {Chen}, \citenamefont {Sank}, \citenamefont {O'Malley},
  \citenamefont {White}, \citenamefont {Barends}, \citenamefont {Kelly},
  \citenamefont {Lucero}, \citenamefont {Mariantoni}, \citenamefont {Megrant},
  \citenamefont {Neill}, \citenamefont {Vainsencher}, \citenamefont {Wenner},
  \citenamefont {Korotkov}, \citenamefont {Cleland},\ and\ \citenamefont
  {Martinis}}]{Yin2013}%
  \BibitemOpen
  \bibfield  {author} {\bibinfo {author} {\bibfnamefont {Y.}~\bibnamefont
  {Yin}}, \bibinfo {author} {\bibfnamefont {Y.}~\bibnamefont {Chen}}, \bibinfo
  {author} {\bibfnamefont {D.}~\bibnamefont {Sank}}, \bibinfo {author}
  {\bibfnamefont {P.~J.~J.}\ \bibnamefont {O'Malley}}, \bibinfo {author}
  {\bibfnamefont {T.~C.}\ \bibnamefont {White}}, \bibinfo {author}
  {\bibfnamefont {R.}~\bibnamefont {Barends}}, \bibinfo {author} {\bibfnamefont
  {J.}~\bibnamefont {Kelly}}, \bibinfo {author} {\bibfnamefont
  {E.}~\bibnamefont {Lucero}}, \bibinfo {author} {\bibfnamefont
  {M.}~\bibnamefont {Mariantoni}}, \bibinfo {author} {\bibfnamefont
  {A.}~\bibnamefont {Megrant}}, \bibinfo {author} {\bibfnamefont
  {C.}~\bibnamefont {Neill}}, \bibinfo {author} {\bibfnamefont
  {A.}~\bibnamefont {Vainsencher}}, \bibinfo {author} {\bibfnamefont
  {J.}~\bibnamefont {Wenner}}, \bibinfo {author} {\bibfnamefont {A.~N.}\
  \bibnamefont {Korotkov}}, \bibinfo {author} {\bibfnamefont {A.~N.}\
  \bibnamefont {Cleland}},\ and\ \bibinfo {author} {\bibfnamefont {J.~M.}\
  \bibnamefont {Martinis}},\ }\href
  {https://doi.org/10.1103/PhysRevLett.110.107001} {\bibfield  {journal}
  {\bibinfo  {journal} {Phys. Rev. Lett.}\ }\textbf {\bibinfo {volume} {110}},\
  \bibinfo {pages} {107001} (\bibinfo {year} {2013})}\BibitemShut {NoStop}%
\bibitem [{\citenamefont {Srinivasan}\ \emph {et~al.}(2014)\citenamefont
  {Srinivasan}, \citenamefont {Sundaresan}, \citenamefont {Sadri},
  \citenamefont {Liu}, \citenamefont {Gambetta}, \citenamefont {Yu},
  \citenamefont {Girvin},\ and\ \citenamefont {Houck}}]{Srinivasan2014}%
  \BibitemOpen
  \bibfield  {author} {\bibinfo {author} {\bibfnamefont {S.~J.}\ \bibnamefont
  {Srinivasan}}, \bibinfo {author} {\bibfnamefont {N.~M.}\ \bibnamefont
  {Sundaresan}}, \bibinfo {author} {\bibfnamefont {D.}~\bibnamefont {Sadri}},
  \bibinfo {author} {\bibfnamefont {Y.}~\bibnamefont {Liu}}, \bibinfo {author}
  {\bibfnamefont {J.~M.}\ \bibnamefont {Gambetta}}, \bibinfo {author}
  {\bibfnamefont {T.}~\bibnamefont {Yu}}, \bibinfo {author} {\bibfnamefont
  {S.~M.}\ \bibnamefont {Girvin}},\ and\ \bibinfo {author} {\bibfnamefont
  {A.~A.}\ \bibnamefont {Houck}},\ }\href
  {https://doi.org/10.1103/PhysRevA.89.033857} {\bibfield  {journal} {\bibinfo
  {journal} {Phys. Rev. A}\ }\textbf {\bibinfo {volume} {89}},\ \bibinfo
  {pages} {033857} (\bibinfo {year} {2014})}\BibitemShut {NoStop}%
\bibitem [{\citenamefont {Zeytinoğlu}\ \emph {et~al.}(2015)\citenamefont
  {Zeytinoğlu}, \citenamefont {Pechal}, \citenamefont {Berger}, \citenamefont
  {Abdumalikov}, \citenamefont {Wallraff},\ and\ \citenamefont
  {Filipp}}]{Zeytinoglu2015}%
  \BibitemOpen
  \bibfield  {author} {\bibinfo {author} {\bibfnamefont {S.}~\bibnamefont
  {Zeytinoğlu}}, \bibinfo {author} {\bibfnamefont {M.}~\bibnamefont {Pechal}},
  \bibinfo {author} {\bibfnamefont {S.}~\bibnamefont {Berger}}, \bibinfo
  {author} {\bibfnamefont {A.~A.}\ \bibnamefont {Abdumalikov}}, \bibinfo
  {author} {\bibfnamefont {A.}~\bibnamefont {Wallraff}},\ and\ \bibinfo
  {author} {\bibfnamefont {S.}~\bibnamefont {Filipp}},\ }\href
  {https://doi.org/10.1103/PhysRevA.91.043846} {\bibfield  {journal} {\bibinfo
  {journal} {Phys. Rev. A}\ }\textbf {\bibinfo {volume} {91}},\ \bibinfo
  {pages} {043846} (\bibinfo {year} {2015})}\BibitemShut {NoStop}%
\bibitem [{\citenamefont {Yang}\ \emph {et~al.}(2006)\citenamefont {Yang},
  \citenamefont {Ya-Ping},\ and\ \citenamefont {Hong}}]{Yang2006}%
  \BibitemOpen
  \bibfield  {author} {\bibinfo {author} {\bibfnamefont {S.}~\bibnamefont
  {Yang}}, \bibinfo {author} {\bibfnamefont {Y.}~\bibnamefont {Ya-Ping}},\ and\
  \bibinfo {author} {\bibfnamefont {C.}~\bibnamefont {Hong}},\ }\href
  {https://doi.org/10.1088/0256-307X/23/5/020} {\bibfield  {journal} {\bibinfo
  {journal} {Chin. Phys. Lett.}\ }\textbf {\bibinfo {volume} {23}},\ \bibinfo
  {pages} {1136} (\bibinfo {year} {2006})}\BibitemShut {NoStop}%
\bibitem [{\citenamefont {Glauber}(1963)}]{Glauber1963c}%
  \BibitemOpen
  \bibfield  {author} {\bibinfo {author} {\bibfnamefont {R.~J.}\ \bibnamefont
  {Glauber}},\ }\href {https://doi.org/10.1103/PhysRev.131.2766} {\bibfield
  {journal} {\bibinfo  {journal} {Phys. Rev.}\ }\textbf {\bibinfo {volume}
  {131}},\ \bibinfo {pages} {2766} (\bibinfo {year} {1963})}\BibitemShut
  {NoStop}%
\bibitem [{\citenamefont {Phoenix}\ and\ \citenamefont
  {Knight}(1988)}]{Phoenix1988}%
  \BibitemOpen
  \bibfield  {author} {\bibinfo {author} {\bibfnamefont {S.}~\bibnamefont
  {Phoenix}}\ and\ \bibinfo {author} {\bibfnamefont {P.}~\bibnamefont
  {Knight}},\ }\href
  {https://doi.org/https://doi.org/10.1016/0003-4916(88)90006-1} {\bibfield
  {journal} {\bibinfo  {journal} {Ann. Phys.}\ }\textbf {\bibinfo {volume}
  {186}},\ \bibinfo {pages} {381} (\bibinfo {year} {1988})}\BibitemShut
  {NoStop}%
\bibitem [{\citenamefont {Birrittella}\ \emph {et~al.}(2015)\citenamefont
  {Birrittella}, \citenamefont {Cheng},\ and\ \citenamefont
  {Gerry}}]{Birrittella2015}%
  \BibitemOpen
  \bibfield  {author} {\bibinfo {author} {\bibfnamefont {R.}~\bibnamefont
  {Birrittella}}, \bibinfo {author} {\bibfnamefont {K.}~\bibnamefont {Cheng}},\
  and\ \bibinfo {author} {\bibfnamefont {C.~C.}\ \bibnamefont {Gerry}},\ }\href
  {https://doi.org/10.1016/j.optcom.2015.05.069} {\bibfield  {journal}
  {\bibinfo  {journal} {Opt. Comm.}\ }\textbf {\bibinfo {volume} {354}},\
  \bibinfo {pages} {286} (\bibinfo {year} {2015})}\BibitemShut {NoStop}%
\bibitem [{\citenamefont {Gerry}\ and\ \citenamefont
  {Mimih}(2010)}]{gerry2010parity}%
  \BibitemOpen
  \bibfield  {author} {\bibinfo {author} {\bibfnamefont {C.~C.}\ \bibnamefont
  {Gerry}}\ and\ \bibinfo {author} {\bibfnamefont {J.}~\bibnamefont {Mimih}},\
  }\href {https://doi.org/https://doi.org/10.1080/00107514.2010.509995}
  {\bibfield  {journal} {\bibinfo  {journal} {Contemp. Phys.}\ }\textbf
  {\bibinfo {volume} {51}},\ \bibinfo {pages} {497} (\bibinfo {year}
  {2010})}\BibitemShut {NoStop}%
\bibitem [{\citenamefont {Scully}\ and\ \citenamefont
  {Zubairy}(1997)}]{Scully1997}%
  \BibitemOpen
  \bibfield  {author} {\bibinfo {author} {\bibfnamefont {M.~O.}\ \bibnamefont
  {Scully}}\ and\ \bibinfo {author} {\bibfnamefont {M.~S.}\ \bibnamefont
  {Zubairy}},\ }\href {https://doi.org/10.1017/CBO9780511813993} {\emph
  {\bibinfo {title} {{Quantum Optics}}}}\ (\bibinfo  {publisher} {Cambridge
  University Press},\ \bibinfo {address} {Cambridge},\ \bibinfo {year}
  {1997})\BibitemShut {NoStop}%
\bibitem [{\citenamefont {Mandel}(1982)}]{Mandel1982}%
  \BibitemOpen
  \bibfield  {author} {\bibinfo {author} {\bibfnamefont {L.}~\bibnamefont
  {Mandel}},\ }\href {https://doi.org/10.1103/PhysRevLett.49.136} {\bibfield
  {journal} {\bibinfo  {journal} {Phys. Rev. Let.}\ }\textbf {\bibinfo {volume}
  {49}},\ \bibinfo {pages} {136} (\bibinfo {year} {1982})}\BibitemShut
  {NoStop}%
\bibitem [{\citenamefont {Meystre}\ and\ \citenamefont
  {Zubairy}(1982)}]{Meystre1982}%
  \BibitemOpen
  \bibfield  {author} {\bibinfo {author} {\bibfnamefont {P.}~\bibnamefont
  {Meystre}}\ and\ \bibinfo {author} {\bibfnamefont {M.}~\bibnamefont
  {Zubairy}},\ }\href {https://doi.org/10.1016/0375-9601(82)90330-9} {\bibfield
   {journal} {\bibinfo  {journal} {Phys. Lett. A}\ }\textbf {\bibinfo {volume}
  {89}},\ \bibinfo {pages} {390} (\bibinfo {year} {1982})}\BibitemShut
  {NoStop}%
\bibitem [{\citenamefont {Knight}(1986)}]{Knight1986}%
  \BibitemOpen
  \bibfield  {author} {\bibinfo {author} {\bibfnamefont {P.~L.}\ \bibnamefont
  {Knight}},\ }\href {https://doi.org/10.1088/0031-8949/1986/T12/007}
  {\bibfield  {journal} {\bibinfo  {journal} {Phys. Scr.}\ }\textbf {\bibinfo
  {volume} {T12}},\ \bibinfo {pages} {51} (\bibinfo {year} {1986})}\BibitemShut
  {NoStop}%
\bibitem [{\citenamefont {Kukli{\'{n}}ski}\ and\ \citenamefont
  {Madajczyk}(1988)}]{Kuklinski1988}%
  \BibitemOpen
  \bibfield  {author} {\bibinfo {author} {\bibfnamefont {J.~R.}\ \bibnamefont
  {Kukli{\'{n}}ski}}\ and\ \bibinfo {author} {\bibfnamefont {J.~L.}\
  \bibnamefont {Madajczyk}},\ }\href {https://doi.org/10.1103/PhysRevA.37.3175}
  {\bibfield  {journal} {\bibinfo  {journal} {Phys. Rev. A}\ }\textbf {\bibinfo
  {volume} {37}},\ \bibinfo {pages} {3175} (\bibinfo {year}
  {1988})}\BibitemShut {NoStop}%
\bibitem [{\citenamefont {Loudon}\ and\ \citenamefont
  {Knight}(1987)}]{Loudon1987}%
  \BibitemOpen
  \bibfield  {author} {\bibinfo {author} {\bibfnamefont {R.}~\bibnamefont
  {Loudon}}\ and\ \bibinfo {author} {\bibfnamefont {P.}~\bibnamefont
  {Knight}},\ }\href {https://doi.org/10.1080/09500348714550721} {\bibfield
  {journal} {\bibinfo  {journal} {J. Mod. Opt.}\ }\textbf {\bibinfo {volume}
  {34}},\ \bibinfo {pages} {709} (\bibinfo {year} {1987})}\BibitemShut
  {NoStop}%
\bibitem [{\citenamefont {Schleich}(2001)}]{Schleich2001}%
  \BibitemOpen
  \bibfield  {author} {\bibinfo {author} {\bibfnamefont {W.~P.}\ \bibnamefont
  {Schleich}},\ }\href {https://doi.org/10.1002/3527602976} {\emph {\bibinfo
  {title} {{Quantum Optics in Phase Space}}}}\ (\bibinfo  {publisher} {Wiley},\
  \bibinfo {address} {Berlin},\ \bibinfo {year} {2001})\BibitemShut {NoStop}%
\bibitem [{\citenamefont {Husimi}(1940)}]{Husimi1940}%
  \BibitemOpen
  \bibfield  {author} {\bibinfo {author} {\bibfnamefont {K.}~\bibnamefont
  {Husimi}},\ }\href {https://doi.org/10.11429/ppmsj1919.22.4_264} {\bibfield
  {journal} {\bibinfo  {journal} {Phys. Math. Soc. Jpn.}\ }\textbf {\bibinfo
  {volume} {22}},\ \bibinfo {pages} {264} (\bibinfo {year} {1940})}\BibitemShut
  {NoStop}%
\bibitem [{\citenamefont {Kano}(1965)}]{Kano1965}%
  \BibitemOpen
  \bibfield  {author} {\bibinfo {author} {\bibfnamefont {Y.}~\bibnamefont
  {Kano}},\ }\href {https://doi.org/10.1063/1.1704739} {\bibfield  {journal}
  {\bibinfo  {journal} {J. Math. Phys.}\ }\textbf {\bibinfo {volume} {6}},\
  \bibinfo {pages} {1913} (\bibinfo {year} {1965})}\BibitemShut {NoStop}%
\bibitem [{\citenamefont {Cahill}\ and\ \citenamefont
  {Glauber}(1969)}]{Cahill1969}%
  \BibitemOpen
  \bibfield  {author} {\bibinfo {author} {\bibfnamefont {K.~E.}\ \bibnamefont
  {Cahill}}\ and\ \bibinfo {author} {\bibfnamefont {R.~J.}\ \bibnamefont
  {Glauber}},\ }\href {https://doi.org/10.1103/PhysRev.177.1882} {\bibfield
  {journal} {\bibinfo  {journal} {Phys. Rev.}\ }\textbf {\bibinfo {volume}
  {177}},\ \bibinfo {pages} {1882} (\bibinfo {year} {1969})}\BibitemShut
  {NoStop}%
\bibitem [{\citenamefont {Cartwright}(1976)}]{Cartwright1976}%
  \BibitemOpen
  \bibfield  {author} {\bibinfo {author} {\bibfnamefont {N.}~\bibnamefont
  {Cartwright}},\ }\href {https://doi.org/10.1016/0378-4371(76)90145-X}
  {\bibfield  {journal} {\bibinfo  {journal} {Phys. A: Stat. Mech. Appl.}\
  }\textbf {\bibinfo {volume} {83}},\ \bibinfo {pages} {210} (\bibinfo {year}
  {1976})}\BibitemShut {NoStop}%
\bibitem [{\citenamefont {Eiselt}\ and\ \citenamefont
  {Risken}(1989)}]{Eiselt1989}%
  \BibitemOpen
  \bibfield  {author} {\bibinfo {author} {\bibfnamefont {J.}~\bibnamefont
  {Eiselt}}\ and\ \bibinfo {author} {\bibfnamefont {H.}~\bibnamefont
  {Risken}},\ }\href {https://doi.org/10.1016/0030-4018(89)90438-0} {\bibfield
  {journal} {\bibinfo  {journal} {Opt. Commun.}\ }\textbf {\bibinfo {volume}
  {72}},\ \bibinfo {pages} {351} (\bibinfo {year} {1989})}\BibitemShut
  {NoStop}%
\bibitem [{\citenamefont {Bu{\v{z}}ek}\ \emph {et~al.}(1992)\citenamefont
  {Bu{\v{z}}ek}, \citenamefont {Moya-Cessa}, \citenamefont {Knight},\ and\
  \citenamefont {Phoenix}}]{Buzek1992}%
  \BibitemOpen
  \bibfield  {author} {\bibinfo {author} {\bibfnamefont {V.}~\bibnamefont
  {Bu{\v{z}}ek}}, \bibinfo {author} {\bibfnamefont {H.}~\bibnamefont
  {Moya-Cessa}}, \bibinfo {author} {\bibfnamefont {P.~L.}\ \bibnamefont
  {Knight}},\ and\ \bibinfo {author} {\bibfnamefont {S.~J.~D.}\ \bibnamefont
  {Phoenix}},\ }\href {https://doi.org/10.1103/PhysRevA.45.8190} {\bibfield
  {journal} {\bibinfo  {journal} {Phys. Rev. A}\ }\textbf {\bibinfo {volume}
  {45}},\ \bibinfo {pages} {8190} (\bibinfo {year} {1992})}\BibitemShut
  {NoStop}%
\bibitem [{\citenamefont {Eiselt}\ and\ \citenamefont
  {Risken}(1991)}]{Eiselt1991}%
  \BibitemOpen
  \bibfield  {author} {\bibinfo {author} {\bibfnamefont {J.}~\bibnamefont
  {Eiselt}}\ and\ \bibinfo {author} {\bibfnamefont {H.}~\bibnamefont
  {Risken}},\ }\href {https://doi.org/10.1103/PhysRevA.43.346} {\bibfield
  {journal} {\bibinfo  {journal} {Phys. Rev. A}\ }\textbf {\bibinfo {volume}
  {43}},\ \bibinfo {pages} {346} (\bibinfo {year} {1991})}\BibitemShut
  {NoStop}%
\bibitem [{\citenamefont {Bu{\v{z}}ek}\ and\ \citenamefont
  {Hladk{\'{y}}}(1993)}]{Buzek1993}%
  \BibitemOpen
  \bibfield  {author} {\bibinfo {author} {\bibfnamefont {V.}~\bibnamefont
  {Bu{\v{z}}ek}}\ and\ \bibinfo {author} {\bibfnamefont {B.}~\bibnamefont
  {Hladk{\'{y}}}},\ }\href {https://doi.org/10.1080/09500349314551371}
  {\bibfield  {journal} {\bibinfo  {journal} {J. Mod. Opt.}\ }\textbf {\bibinfo
  {volume} {40}},\ \bibinfo {pages} {1309} (\bibinfo {year}
  {1993})}\BibitemShut {NoStop}%
\bibitem [{\citenamefont {Alsing}\ and\ \citenamefont
  {Carmichael}(1991)}]{Alsing1991}%
  \BibitemOpen
  \bibfield  {author} {\bibinfo {author} {\bibfnamefont {P.}~\bibnamefont
  {Alsing}}\ and\ \bibinfo {author} {\bibfnamefont {H.~J.}\ \bibnamefont
  {Carmichael}},\ }\href {https://doi.org/10.1088/0954-8998/3/1/003} {\bibfield
   {journal} {\bibinfo  {journal} {J. Eur. Opt. Soc. Part B}\ }\textbf
  {\bibinfo {volume} {3}},\ \bibinfo {pages} {13} (\bibinfo {year}
  {1991})}\BibitemShut {NoStop}%
\bibitem [{\citenamefont {{M. Fang}}(1998)}]{Fang1998}%
  \BibitemOpen
  \bibfield  {author} {\bibinfo {author} {\bibnamefont {{M. Fang}}},\ }\href
  {https://doi.org/https://doi.org/10.1016/S0378-4371(98)00234-9} {\bibfield
  {journal} {\bibinfo  {journal} {Phys. A: Stat. Mech. Appl.}\ }\textbf
  {\bibinfo {volume} {259}},\ \bibinfo {pages} {193} (\bibinfo {year}
  {1998})}\BibitemShut {NoStop}%
\bibitem [{\citenamefont {Phoenix}\ and\ \citenamefont
  {Knight}(1991)}]{Phoenix1991}%
  \BibitemOpen
  \bibfield  {author} {\bibinfo {author} {\bibfnamefont {S.~J.~D.}\
  \bibnamefont {Phoenix}}\ and\ \bibinfo {author} {\bibfnamefont {P.~L.}\
  \bibnamefont {Knight}},\ }\href {https://doi.org/10.1103/PhysRevA.44.6023}
  {\bibfield  {journal} {\bibinfo  {journal} {Phys. Rev. A}\ }\textbf {\bibinfo
  {volume} {44}},\ \bibinfo {pages} {6023} (\bibinfo {year}
  {1991})}\BibitemShut {NoStop}%
\bibitem [{\citenamefont {Shore}\ and\ \citenamefont
  {Knight}(1993)}]{SHORE1993}%
  \BibitemOpen
  \bibfield  {author} {\bibinfo {author} {\bibfnamefont {B.~W.}\ \bibnamefont
  {Shore}}\ and\ \bibinfo {author} {\bibfnamefont {P.~L.}\ \bibnamefont
  {Knight}},\ }\href {https://doi.org/10.1080/09500349314551321} {\bibfield
  {journal} {\bibinfo  {journal} {J. Mod. Opt.}\ }\textbf {\bibinfo {volume}
  {40}},\ \bibinfo {pages} {1195} (\bibinfo {year} {1993})}\BibitemShut
  {NoStop}%
\end{thebibliography}

\end{document}